\documentclass[preprint,12pt,authoryear]{elsarticle}

\usepackage[a4paper,total={6.5in,9.5in}]{geometry}
\usepackage{amssymb}
\usepackage{amsmath}
\usepackage{xcolor}
\usepackage{lineno,hyperref}
\usepackage{subcaption}
\usepackage{algorithm}
\usepackage{algpseudocode}
\usepackage{amsthm}

\usepackage{lineno}

\begin{document}

\begin{frontmatter}

\title{
Efficacy of the Weak Formulation of Sparse Nonlinear Identification in Predicting Vortex-Induced Vibrations
} 

\author[1]{Haimi Jha}

\address[1]{Department of Civil Engineering, Jadavpur University, Kolkata, India}

\author[2]{Hibah Saddal}


\address[2]{Aerospace Engineering, School of Metallurgy and Materials, University of Birmingham, Birmingham, UK}

\author[2]{Chandan Bose}
\ead{c.bose@bham.ac.uk}
\cortext[cor1]{Corresponding author}

\begin{abstract}
Vortex-induced vibrations (VIV) remain a canonical yet complex manifestation of fluid-structure interactions, where coupled nonlinear dynamics govern the motion of bluff bodies. For several years, we have relied on traditional reduced-order mathematical models derived from empirical and oscillator-based formulations; however, such models often fail to reproduce the quantitative dynamics observed in realistic flow environments. In this study, we explore a data-driven framework that leverages sparse identification of nonlinear dynamics (SINDy) and its weak formulation to uncover the governing equations of a single-degree-of-freedom cylinder undergoing VIV, using both data generated from previously developed reduced-order models and high-fidelity simulation results to assess the interpretation and efficacy of models discovered from a purely data-driven approach, particularly when the underlying dynamics are not fully known. The weak formulation (WSINDy), which replaces numerical differentiation with an integral-based representation, demonstrates marked robustness for aperiodic dynamics in particular. A complementary analysis using proper orthogonal decomposition (POD) is employed to extract the dominant spatio-temporal structures of the flow and to assess whether the temporal evolution of the wake can be represented on a reduced-dimensional manifold. The findings establish that data-driven identification can recover interpretable, quantitatively reliable models of VIV, providing a robust and computationally efficient pathway for modelling fluid--structure interactions directly from data. In particular, WSINDy is shown to be a more robust and interpretable alternative to standard SINDy for discovering VIV equations from aperiodic response dynamics, paving the way for predictive, data-informed design of fluid--structure interaction systems.
\end{abstract}




\end{frontmatter}



\section{Introduction}
When flexible structures are subjected to unsteady fluid flow, the flow-induced forces may often lead to self-sustained vibrations of the structure beyond a critical non-dimensional flow velocity, a phenomenon known as flow-induced vibration (FIV). Vortex-induced vibration (VIV) is a distinct subclass of FIV, characterised by a specific frequency lock-in regime, and is often encountered in cylindrical structures with circular cross-sections, such as marine risers, chimneys, and bridge cables~\citep{Williamson_VIV,bearman2011circular,annurev:/content/journals/10.1146/annurev-fluid-022724-014235}. Flow around bluff bodies leads to vortex shedding due to flow separation, and the interaction of these opposite-signed vortices, shed in an alternate fashion, with the structure generates oscillatory forces, which induce oscillations in the flexible structures along their degrees-of-freedom. Predicting VIV response is crucial for design, as prolonged exposure to such oscillations can lead to fatigue damage, thereby affecting structural integrity. Conversely, when properly understood and controlled, VIV can also be exploited for harnessing wind or wave energy.

Developing low-order mathematical models to predict and control the complex behaviour of VIV has attracted significant research attention in the past. To that end, one widely adopted approach is to model the unsteady loads in terms of a wake parameter, modelled as a non-linear Van der Pol oscillator, commonly referred to as the wake oscillator~\citep{JapanArchitecturalReviewPaper,Facchinetti}. In this approach, a mass-spring-damper oscillator equation describing the structure's motion is coupled with the wake oscillator equation. Different coupling terms, such as displacement, velocity, or acceleration coupling, can be employed to model the interaction between the structure and the wake~\citep{Facchinetti}. It is worth noting that these models impose a limit-cycle topology on the wake dynamics that is inherently two-dimensional and single-frequency. In reality, the near-wake of a bluff body undergoing VIV exhibits complex three-dimensional vortex dynamics, multi-frequency spectral content, and mode competition; phenomena such as the 2P and 2S shedding modes identified by \citet{Williamson_VIV} lie entirely beyond the representational capacity of the  Van der Pol equation. Furthermore, the empirical nonlinear parameters and the coupling coefficient must be calibrated against experimental data and have been shown to lack universality across different mass-damping parameters, aspect ratios, and Reynolds number regimes \citep{Facchinetti}, severely limiting the model's generalisability. 

These phenomenological reduced-order mathematical models rely on empirical fitting, grounded in physical intuition and understanding. However, such approaches may overlook critical terms necessary to accurately capture the essential non-linear dynamics. Recent work on data‐driven discovery of dynamical systems has made it increasingly plausible to move toward interpretable governing equations inferred directly from measured or simulated time series. In this context, the Sparse Identification of Nonlinear Dynamics (SINDy) framework, developed by \citet{Brunton_2016}, enables the discovery of governing equations for dynamical systems directly from time-series data. By performing sparse regression over a library of candidate functions, it seeks the parsimonious set of terms that best explains the observed dynamics. 
In this way, SINDy provides a simplified yet accurate representation of nonlinear behaviour and offers a promising alternative for directly extracting governing equations from simulated or experimental data.

A key practical limitation of standard SINDy in fluid–structure interaction and VIV applications is its strong dependence on pointwise time derivatives of measured signals, which are very sensitive to noise and numerical differentiation errors. In VIV problems, force and displacement time series are often perturbed by measurement noise, broadband wake fluctuations, and multiscale dynamics, making reliable derivative estimation a dominant source of error and frequently leading to spurious or nonphysical terms in the identified model. Weak SINDy (WSINDy) was developed in 2020 by \citet{weakSINDy} to address this limitation by reformulating the identification problem in a weak (integral) sense, replacing pointwise derivatives with weighted time integrals with respect to smooth test functions. For VIV, where experimental and high-fidelity computational fluid dynamics (CFD) data are often limited, noisy, and not amenable to robust differentiation, WSINDy therefore provides a crucial methodological advancement, enabling sparse, interpretable discovery of governing equations while preserving physical coupling mechanisms, without requiring ad hoc smoothing or derivative tuning.

For predicting VIV response, where nonlinearity, mode coupling, and regime changes are instrumental, this approach is compelling; however, current efforts to understand VIV dynamics using SINDy remain largely unexplored, except for a few recent studies. \citet{HONI2022} demonstrated the use of sparse regression to uncover hidden coupling terms in reduced-order models of VIV. A synthetic dataset was generated by numerically integrating a modified wake oscillator system that included a non-linear forcing term, 
which was not known to the identification algorithm. By employing different function libraries, including a physically informed one, the study showed that SINDy could infer this hidden term solely from the observed time histories of displacement and velocity. A comparison between the generic and physically informed libraries further emphasised SINDy's ability to accept structurally different yet functionally equivalent terms, underscoring its potential to identify coupling dynamics in reduced-order models. A limitation, however, is that the study focuses exclusively on coupling terms derived from numerical simulations, without extending the framework to full-system modelling.

\citet{LELKES2023117847} applied a similar sparse modelling framework to construct reduced-order models of aeroelastic systems. The study considered a flat plate undergoing large-amplitude heaving and pitching motions, with state variables including plunge velocity, plunge acceleration, pitch angle, and pitch rate.
These variables served as basis functions for identifying the governing equations for the time derivatives of the aerodynamic force and moment coefficients, which describe the unsteady aerodynamic response to plunge and pitch. A polynomial representation accurately captured the complex aerodynamic interactions, and the identified models showed close agreement with CFD results across a range of amplitudes and frequencies. The authors noted, however, that further experimental validation is needed to fully assess the framework's robustness.

\citet{cheng2024unifiedframeworkpredictionvortexinduced} investigated a grey-box identification strategy in which the governing equation structure is prescribed a priori as a general nonlinear oscillator, whilst unknown coefficients governing nonlinear damping and fluid-structure energy exchange are identified from data, demonstrating physically interpretable models with robust performance at $Re=100$. However, this approach is fundamentally constrained by its dependence on prior structural knowledge, limiting its applicability to VIV regimes where dominant nonlinearities, such as nonlinear fluid damping, amplitude-dependent added mass, or higher-harmonic wake contributions, are not well characterised. \citet{fukami2021sparse} coupled a CNN autoencoder with SINDy to identify governing equations without a priori assumptions, but identification is performed in an abstract latent space rather than the physical state space of the coupled fluid-structure system. All demonstrations are confined to a stationary, rigidly fixed cylinder with no fluid-structure coupling, and validation is limited to reproducing in-sample trajectories. Overall, grey-box approaches preserve interpretability but require a prescribed structure, whilst latent-space sparse identification removes this constraint at the cost of physical transparency and coupled-system fidelity. The present study addresses both limitations by applying SINDy and WSINDy directly to the physical state space of the coupled VIV system.

Existing studies indicate that while sparse regression and SINDy-based approaches \citep{Brunton_2016} have advanced the identification of reduced-order representations for VIV-related dynamics, their application remains largely confined to controlled or partially prescribed settings. Most demonstrations focus on recovering missing or hidden terms within assumed wake-oscillator frameworks \citep{HONI2022}, fitting compact surrogates for unsteady loads \citep{fukami2021sparse}, or calibrating grey-box models with a priori-specified structure \citep{cheng2024unifiedframeworkpredictionvortexinduced}. As a result, a clear gap persists between term discovery and the more challenging task of governing equation discovery for VIV when the underlying dynamics are inferred directly from CFD or experimental data, where forcing is broadband, noise-contaminated, and strongly regime-dependent \citep{li2019discovering}. Moreover, existing formulations often treat the wake and the structure separately or model only part of the coupled system, thereby limiting insight into full-system identifiability and energy-exchange mechanisms. The non-uniqueness of sparse representations further complicates this problem, as generic libraries can yield structurally distinct yet functionally equivalent models that fit data but violate physical constraints \citep{loiseau2018constrained}. Additional gaps arise in the robustness of SINDy models across canonical VIV regimes, including lock-in, desynchronisation, and hysteresis, as well as in handling noise and in derivative estimation in realistic datasets, where differentiation errors frequently dominate \citep{kaheman2020sindy, fasel2022ensemble, weakSINDy}. Finally, validation has largely focused on trajectory agreement rather than on VIV-specific predictive metrics, such as amplitude response curves, phase relationships, etc \citep{li2019discovering}. Collectively, these limitations highlight the need for SINDy frameworks tailored to VIV that can discover coupled governing equations in aperiodic or noisy states, embed physical constraints, and demonstrate predictive robustness across regimes.

The present study aims to address these gaps by conducting the first systematic comparative assessment of standard SINDy and its weak formulation, WSINDy, for governing equation discovery in vortex-induced vibration. Particular emphasis is placed on examining the predictive performance of both approaches across all canonical response regimes, including the aperiodic regime in which conventional strong-form sparse regression is shown to deteriorate. Standard SINDy constructs the sparse regression problem directly from numerically differentiated state trajectories, rendering it inherently susceptible to noise amplification and sampling irregularities. WSINDy, by contrast, projects the identification problem onto a space of compactly supported test functions, converting derivative operations into weighted integrals that act as a natural low-pass filter on the data \citep{weakSINDy}. This reformulation is particularly consequential for VIV, where the transition from periodic lock-in to aperiodic or quasi-periodic desynchronisation produces broadband spectral content and rapid amplitude modulation, rendering finite-difference derivative estimates unreliable. By directly contrasting the two formulations under identical library, data, and noise conditions, the present work provides a rigorous and controlled basis for quantifying the extent to which weak-form integration mitigates derivative-estimation error and improves the physical consistency of discovered governing equations across the full reduced-velocity range, including the hysteretic lock-in boundary and the aperiodic pre- and post-lock-in regimes.

Another important novelty of this study is full-system coupled identification and VIV-specific validation, both of which are largely absent from the sparse regression literature. Existing SINDy-based VIV studies have either treated the wake and structural subsystems separately, recovered only missing terms within an assumed wake-oscillator framework, or confined validation to trajectory agreement under in-sample conditions, thereby fundamentally precluding confident use of discovered models as predictive tools. The present work closes this gap through a two-stage identification campaign that imposes no a priori structural assumption on the governing equations. In the first stage, synthetic time-series data are generated by numerically integrating the coupled wake oscillator model across a range of reduced velocities, providing a controlled environment in which the ground-truth equations are known, and the full-system identifiability of the coupled structural-wake dynamics can be assessed rigorously. In the second stage, both SINDy and WSINDy are applied directly to high-fidelity CFD data obtained from high-fidelity simulations of an elastically mounted 1-DoF circular cylinder, where the time-series data represent different dynamical regimes, with no prescribed structural assumptions. Critically, model fidelity is evaluated not only by trajectory matching but also by VIV-specific predictive metrics, including amplitude response curves and reconstructed flow fields using the high-energy spatial modes obtained from proper orthogonal decomposition across different reduced-velocity conditions, thereby assessing whether the discovered models capture the underlying physics rather than merely interpolating the training data. 

The primary objectives of this study are as follows: (i) to conduct the first systematic comparative assessment of standard SINDy and WSINDy for governing equation discovery in VIV, directly contrasting both formulations under identical library and data across the full reduced-velocity range; (ii) to perform full-system coupled identification of the VIV response, without imposing any a priori assumption on governing equation structure, and assess model identifiability and sparse regression performance on synthetic time-series data generated from the coupled wake oscillator and high-fidelity CFD data from simulations of an elastically mounted 1-DoF circular cylinder; and (iii) to evaluate the extent to which the weak-form integral projection of WSINDy mitigates derivative-estimation error and improves the physical consistency of discovered governing equations, particularly in the aperiodic desynchronisation regimes where strong-form differentiation is shown to deteriorate.

The remainder of this paper is structured as follows. Sec.~\ref{sec:2} presents the mathematical formulation of the standard SINDy and WSINDy frameworks, including the sparse regression procedure, library construction, and weak-form integral projection. Sec.~\ref{sec:3} applies the regular SINDy frameworks to synthetic time-series data generated from the coupled wake oscillator model across a range of reduced velocities, providing a controlled benchmark for assessing full-system identifiability. Sec.~\ref{sec:4} extends the identification campaign to high-fidelity CFD data obtained from simulations of an elastically mounted 1-DoF circular cylinder, evaluating robustness under broadband, realistic data conditions. Sec.~\ref{sec:5} further investigates SINDy-based identification of reconstructed flow-field dynamics using Proper Orthogonal Decomposition, exploring the extent to which sparse regression can recover interpretable reduced-order representations from modal coefficient data. Sec.~\ref{sec:6} discusses the physical interpretability and structural consistency of identified sparse models. Finally, Sec.~\ref{sec:7} summarises the key findings, draws conclusions on the comparative performance of SINDy and WSINDy for VIV identification, and outlines directions for future work. Supporting material is provided in four appendices. \ref{app:A} details the coupled wake oscillator model used to generate synthetic training data, including the governing equations and numerical integration procedure. \ref{app:B} describes the CFD methodology employed in the high-fidelity simulations, covering mesh generation, boundary conditions, and solver settings. \ref{app:C} presents the Proper Orthogonal Decomposition analysis of the flow-field data, including the modal energy distribution and reconstruction accuracy. \ref{app:D} provides a representative example of the governing equations identified by both SINDy and WSINDy at a reduced velocity of $5$, illustrating the structure and physical consistency of the discovered models.

\section{Formulation of sparse identification of nonlinear dynamics framework}
\label{sec:2}

\noindent The SINDy framework, proposed in 2016 by \citet{Brunton_2016}, is used to discover interpretable, parsimonious equations that model the system's evolution directly from measured data. SINDy aims to accurately capture the system's time evolution using a prescribed set of state variables. Thus, the dimensionality must be predetermined by the number and type of variables used. It is also crucial to have prior knowledge of the system's behaviour before applying this framework. In this study, we leverage prior knowledge from the low-order phenomenological wake-oscillator model \citep{Facchinetti} widely used to predict the VIV response of bluff bodies. In this regard, we will consider the transverse displacement of the cylinder and a wake parameter, as discussed in \ref{app:A}, to serve as the basis for identifying the governing dynamics. Additionally, to reconstruct the fluid flow in reduced dimensions, a different set of state variables, comprising temporal coefficients, is used to govern the evolution of the dominant spatial modes corresponding to the dominant coherent structures, as obtained from Proper Orthogonal Decomposition (POD).

In either approach, once the relevant state variables are defined, the dynamics of the selected state vector $X(t) \in \mathbb{R}^n$ is assumed to follow an ordinary differential equation:
\begin{equation}
\dot{X}(t) = f(X(t)),
\end{equation}
\noindent
where \( f \) is an unknown function to be approximated as a sparse linear combination of candidate nonlinear functions. These candidate functions are organised into a library \( \Phi(X) \), typically consisting of polynomials, trigonometric terms, or other physically motivated functions. For example, a polynomial library up to degree four can be written as:

\begin{equation}
\label{10}
\Phi(X) = \left[1, \, y_1, \, y_2, \, q_1, \, q_2, \, y_1^2, \, y_1 y_2, \, y_1 q_1, \, \ldots, \, q_2^4\right]^T.
\end{equation}

\noindent Here, \( \Phi(X) \) represents the matrix of polynomial basis functions up to the fourth degree. An alternative is to construct a custom library of candidate functions that capture the evolution of state variables.
In this study, we explore two formulations of SINDy: the conventional (regular) SINDy formulation and the WSINDy formulation, which is more robust to noisy and stiff systems. 

\subsection{Regular Formulation of SINDy}
\label{sec:2.1}
\noindent The conventional SINDy formulation presented in \citet{Brunton_2016} relies on the state variable $X(t)$, with its time derivative $\dot{X}(t)$ computed internally via numerical methods, such as finite differences or smoothed finite differences. After estimation of derivatives, relevant basis functions are assembled into the $\Phi(X)$ library. Following this, the sparse regression problem, as given by Eqs.~\ref{29} and \ref{30}, is solved, and a system of governing ODEs for the state variables is identified. A major complication arises in numerically estimating derivatives, as the process is highly sensitive to noise. In practice, even a small amount of noise in \( X(t) \) can lead to large errors in $\dot{X}(t)$, which then propagate into the identified model. This is a significant limitation when applying the framework to experimental or high-fidelity simulation data, where the true governing equations are unknown. This drawback motivated the development of the weak formulation of SINDy \citep{weakSINDy}, which is discussed in the following subsection. 

The regular SINDy model in this study uses the Sequential Thresholded Least Squares (STLSQ) optimiser to identify the set of sparse terms that best fit the data. The optimisation problem can be formulated as:

\begin{equation}
\label{29}
\dot{X} = \Phi(X) \Xi,
\end{equation}

\noindent where $\dot{X}$ is the time derivative of the state variables, $\Phi(X)$ is the basis function matrix, and $\Xi$ is the matrix of coefficients identified for the terms of the basis function. The objective is to find $\Xi$ such that the following optimisation problem is solved:

\begin{equation}
\label{30}
\min_{\Xi} \| \dot{X} - \Phi(X) \Xi \|_2^2 \quad \text{subject to} \quad \| \Xi \|_0 \leq s_t,
\end{equation}

\noindent where \( \| \cdot \|_2^2 \) is the squared \( L_2 \) norm, \( \| \cdot \|_0 \) is the \( L_0 \) norm (which counts the number of nonzero entries), and \( s_t \) is the sparsity threshold.

The thresholding step retains a subset of sparse terms whose coefficients must lie below a specified threshold. Note that the application of the optimiser and the regulariser differs. Optimisation refers to the algorithmic procedure used to solve the regression problem, i.e., to determine the coefficients of $\Xi$. Regularisation refers to additional penalties or constraints that bias the solution toward sparsity or stability (e.g., \(L_1\) and \(L_2\) penalties). STLSQ solves an ordinary least-squares regression to obtain dense coefficients in \(\Xi\) and then applies a hard thresholding step to zero out coefficients whose magnitudes fall below a user-defined threshold. The regression is then restricted to the active terms until convergence. This is computationally efficient and quick. STLSQ does not employ any explicit regularisation methods. Instead, it uses an implicit sparsity-promoting mechanism that approximates the \(L_0\) constraint as in Eq.~\ref{30}. Alternative approaches in literature include the Least Absolute Shrinkage and Selection Operator (LASSO) \citep{tibshirani1996regression}, which adds an \(L_1\) penalty on coefficients; ridge regression \citep{marquardt1975ridge}, which adds an \(L_2\) penalty; and Elastic Net \citep{zou2005regularization}, which combines both. Additionally, methods such as Sequential Thresholded Ridge Regression \citep{zhang2022ridge} exist. Despite these alternatives, this study employs STLSQ, consistent with the original SINDy formulation by \citet{Brunton_2016}, due to its simplicity and ease of implementation. 

\begin{algorithm} [htbp]
\caption{Workflow for Sparse Discovery of Governing Equations using SINDy and WSINDy}
\begin{algorithmic}[1]

\State \textbf{Prepare the data:}
\begin{itemize}
    \item Collect time-series data $X(t)$ from the system of interest.
    \item Apply preprocessing steps such as standardisation or smoothing, 
    especially if the data are noisy.
\end{itemize}

\State \textbf{Choose the feature library:}
\begin{itemize}
    \item Construct a set of candidate nonlinear functions to represent the 
    dynamics. This library may be based on polynomials of a chosen degree, 
    Fourier modes, or a custom-designed set of terms.
\end{itemize}

\State \textbf{Set up the model:}
\begin{itemize}
    \item Decide between using standard SINDy or WSINDy.
    \item Select an appropriate optimiser such as Sequential Thresholded 
    Least Squares (STLSQ).
    \item For WSINDy, define the number, shape, and placement of test 
    functions $\phi(t)$. In this study, we explored evenly spaced test 
    functions with compact support.
\end{itemize}

\For{each scenario (e.g., reduced velocity, flow regime, or configuration)}
    \State Fit the model to the available data $X(t)$ to identify the 
    governing equations.
    
    \State For standard SINDy, compute:
    \begin{equation}
        \dot{X}(t) \approx \Phi(X(t))\, \Xi.
    \end{equation}
    
    \State For WSINDy, minimise the integrated residual:
    \begin{equation}
        R(\mathbf{w}; \phi) = \int_a^b \left[ \phi'(t) \, X(t) 
        + \phi(t) \sum_{j=1}^J w_j f_j(X(t)) \right] dt.
    \end{equation}
    
    \State Refine the model by pruning spurious or noise-sensitive terms as 
    needed. In some cases, certain terms may be removed due to concerns about 
    stability or interpretability.
    
    \State Simulate the discovered model forward in time and store the 
    resulting trajectories.
\EndFor

\State \textbf{Validate the results:}
\begin{itemize}
    \item Compare the discovered model output with reference data through 
    plots of time series, dominant frequency spectra, and structural amplitude.
    \item For CFD-derived data, validation may also include comparison of 
    trajectories, reconstructed fields, etc.
\end{itemize}

\end{algorithmic}
\end{algorithm}

\subsection{Weak Formulation of SINDy}
\noindent WSINDy was developed by \citet{weakSINDy} to directly address the derivative-estimation sensitivity of standard SINDy identified in Sec.~\ref{sec:2.1}, wherein pointwise numerical differentiation of noisy state trajectories yields inaccurate and unstable model discovery. Rather than computing derivatives explicitly, WSINDy reformulates the identification problem in terms of a weak integral form of the governing equations, converting derivative operations into weighted integrals that act as a natural low-pass filter on the data. 

The weak formulation multiplies the governing equations by a set of smooth test functions $\phi(t)$ and integrates over time, yielding residuals of the form:

\begin{equation}
    R(\mathbf{w}; \phi) = \int_a^b \left[ \phi'(t) \, X(t) + \phi(t) \sum_{j=1}^J w_j f_j(X(t)) \right] dt,
\end{equation}
where $X(t)$ is the state vector, \(f_j(X(t))\) are candidate library functions, and \(\mathbf{w}\) is the coefficient vector to be identified. In this context, test functions are compactly supported, smooth functions, such as bump functions or piecewise polynomials, that probe the dynamics locally, thereby avoiding direct differentiation and filtering high-frequency noise. The residual is minimised over a set of test functions, thus improving performance as the number of integration points per subdomain increases. In our implementation, we employed uniformly spaced test functions over the temporal domain. While the original study~\citep{weakSINDy} demonstrates that placing test functions near steep gradients improves coefficient recovery, we opted for uniform spacing for its simplicity and adequate performance in the considered regimes. An important difference lies in the interpretation of the model's output, as in the weak form, {\tt model.predict} returns a weak form of \(\dot{X}\), not the pointwise derivative. This can be unintuitive, especially when compared to the standard SINDy output. One simple workaround is to impose the WSINDy-coefficients on a regular SINDy model and use it to predict $\dot{X}$. Some caveats include the need for identical candidate libraries and their ordering, and identical spatial grids when using PDEs. 

In this WSINDy framework, we employed the Sparse Relaxed Regularised Regression (SR3) optimiser to solve the underlying sparse regression problem, introduced by \citet{zheng2018unifiedframeworksparserelaxed}. It extends sparse regression by introducing a relaxation variable. The relaxation variable, combined with least-squares fitting, improves stability in noisy, ill-conditioned problems. It controls the strength of the coupling between the regression fit and the sparsity constraint; smaller values enforce stronger coupling, yielding a tighter fit, whereas larger values relax this coupling, allowing a smoother transition between active and inactive coefficients. In WSINDy, the regression matrices are formed by integrating states against test functions. Although this reduces noise, it can also lead to ill-conditioned regression problems, particularly when test functions overlap. Thus, an explicit regularisation is needed. SR3 incorporates an explicit regularisation penalty that controls sparsity, with its relaxation variable helping to balance accuracy and sparsity, thereby improving the recovery of the correct terms in the governing equations. Notably, \citet{weakSINDy} also employed SR3 in their original WSINDy formulation for precisely this reason.

Mathematically, SR3 introduces a relaxation variable $\mathbf{w}$ to decouple the regression fit from the sparsity constraint. In the context of SINDy, the optimisation problem is written as:

\begin{equation}
\min_{\Xi, \mathbf{w}} \; \frac{1}{2} \| \dot{X} - \Phi(X)\Xi \|_2^2 + \lambda R(\mathbf{w}) + \frac{\kappa}{2} \| \Xi - \mathbf{w} \|_2^2,
\end{equation}

\noindent where $\dot{X}$ is the time derivative of the state vector, $\Phi(X)$ is the candidate function library, $\Xi$ is the coefficient matrix to be identified, $\mathbf{w}$ is the relaxed variable, $R(\mathbf{w})$ is a sparsity-promoting penalty (e.g., $\ell_1$ norm, hard thresholding, or other regularisers), $\lambda$ is the regularisation weight controlling sparsity, and $\kappa > 0$ enforces closeness between $\Xi$ and $\mathbf{w}$. SR3 is more computationally expensive than STLSQ, as it requires iterative updates of both $\Xi$ and $\mathbf{w}$. The workflow for sparse discovery of governing equations using SINDy and WSINDy is presented in Algorithm 1.




\section{VIV system identification from low-order wake-oscillator models}
\label{sec:3}

\begin{figure*}[htbp]
    \centering
    \includegraphics[width=0.9\textwidth]{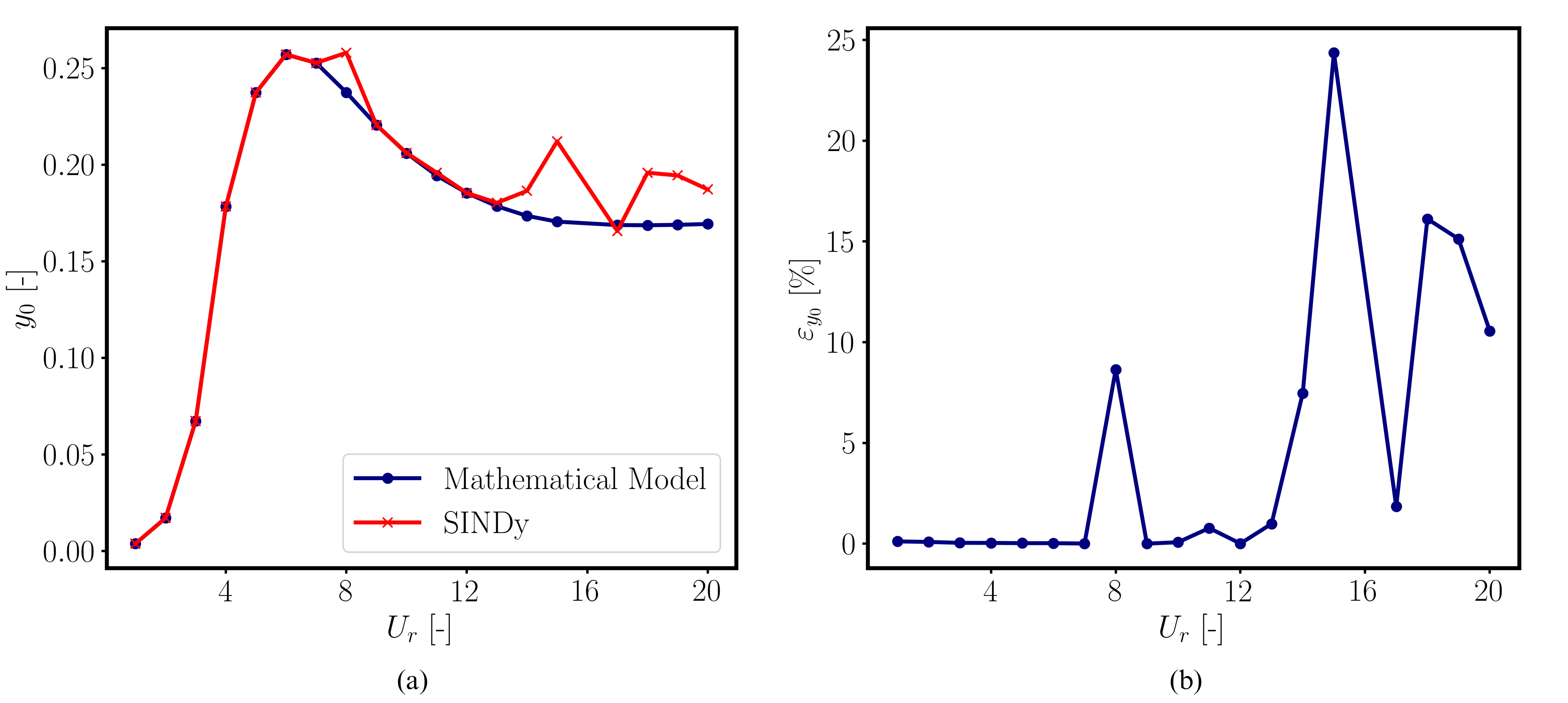}
    \caption{(a) Comparison between SINDy-predicted and numerically integrated non-dimensional structural amplitude ($y_0=\frac{A_y}{D}$, where $A_y$ is the dimensional amplitude) values obtained from the wake-oscillator model for varying reduced velocity $U_r$; (b) percentage relative error ($\epsilon_{y_0}$) in the SINDy-predicted $y_0$ for different $U_r$ values.}
    \label{fig:errorplot}
\end{figure*}

\noindent This section first presents the application of standard SINDy to time-series data generated from the coupled wake oscillator model described in \ref{app:A}, with the objective of assessing whether sparse regression can autonomously recover the underlying governing equations of the coupled structural-wake system without any a priori assumption on their structure. The numerical solutions of Eqs.~\eqref{16}--\eqref{19} for the state variables \( y_1 \), \( y_2 \), \( q_1 \), and \( q_2 \) are computed across a range of reduced velocities \( U_r (=\frac{U_\infty}{f_n D}) \in [1, 20] \), encompassing the pre-lock-in, lock-in, and post-lock-in regimes, and the resulting time-series data are used as input to the identification algorithm. 
Here, $U_\infty$ is the free-stream flow velocity, $f_n$ is the structural natural frequency and $D$ is the diameter of the cylinder.
Since the ground-truth governing equations are known in this setting, the synthetic wake-oscillator data provide a controlled and rigorous benchmark for evaluating the identifiability of the full coupled system, the sparsity and physical consistency of the discovered models, and the comparative performance of standard and weak-form sparse regression prior to application to the more challenging high-fidelity CFD data in Sec.~\ref{sec:4}. The basis function matrix was created using the \texttt{PolynomialFeatures} function. A polynomial library of degree 4 was utilised here, even though prior knowledge of the governing dynamics suggested no degree 4 terms, to test the model's capacity to identify correct dynamics when the candidate library is larger than necessary. The polynomial library constructs nonlinear features that form candidate terms in the governing equations. While a higher degree may offer more expressive power, it can also lead to instability or overfitting, especially at higher \( U_r \). Figure~\ref{fig:errorplot}(a) presents a comparison of the predicted non-dimensional VIV response amplitude $y_0$ obtained from SINDy and direct numerical integration of Eqs.~\eqref{16}--\eqref{19} across the full range of reduced velocities $U_r \in [1, 20]$. The SINDy-identified models demonstrate good agreement with the numerical reference across the majority of the reduced-velocity range, successfully reproducing the onset and sustained lock-in behaviour observed under acceleration coupling, consistent with the wake-oscillator response shown in Fig.~\ref{figure: plotURy01}. Figure~\ref{fig:errorplot}(b) presents the corresponding percentage relative error in the SINDy-predicted structural amplitude, which remains acceptably low throughout the pre-lock-in and lock-in regimes but increases notably at higher reduced velocities, particularly for $U_r \geq 15$. This degradation in predictive accuracy at elevated $U_r$ is attributable to changes in the system dynamics in the post-lock-in regime, where stronger amplitude modulation and richer spectral content are observed. As the dynamics vary with $U_r$, the relative scaling of the coefficients may shift, thereby influencing the retention of certain terms during model identification and contributing to the observed discrepancies.
\begin{figure*}[htbp]
    \centering
    \includegraphics[width=0.8\textwidth]{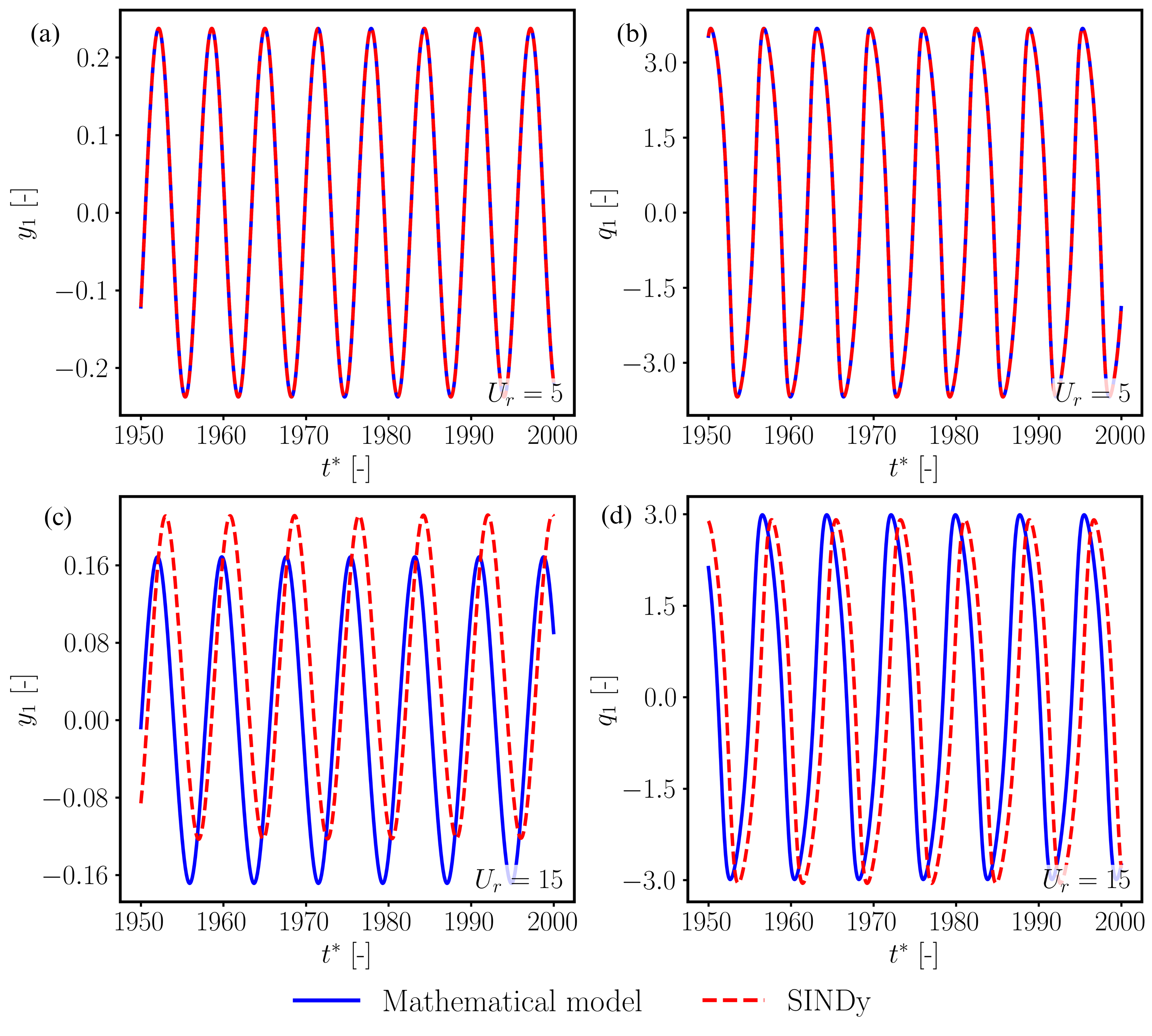}
    \caption{Time histories of (a) $y_1$ and (b) $q_1$ for $U_r = 5$; and (c) $y_1$ and (d) $q_1$ for $U_r = 15$. A shorter time range is plotted to more clearly highlight the dynamics.}
    \label{fig:y1q1short}
\end{figure*}
\begin{figure*}[htbp]
    \centering
    \includegraphics[width=1\textwidth]{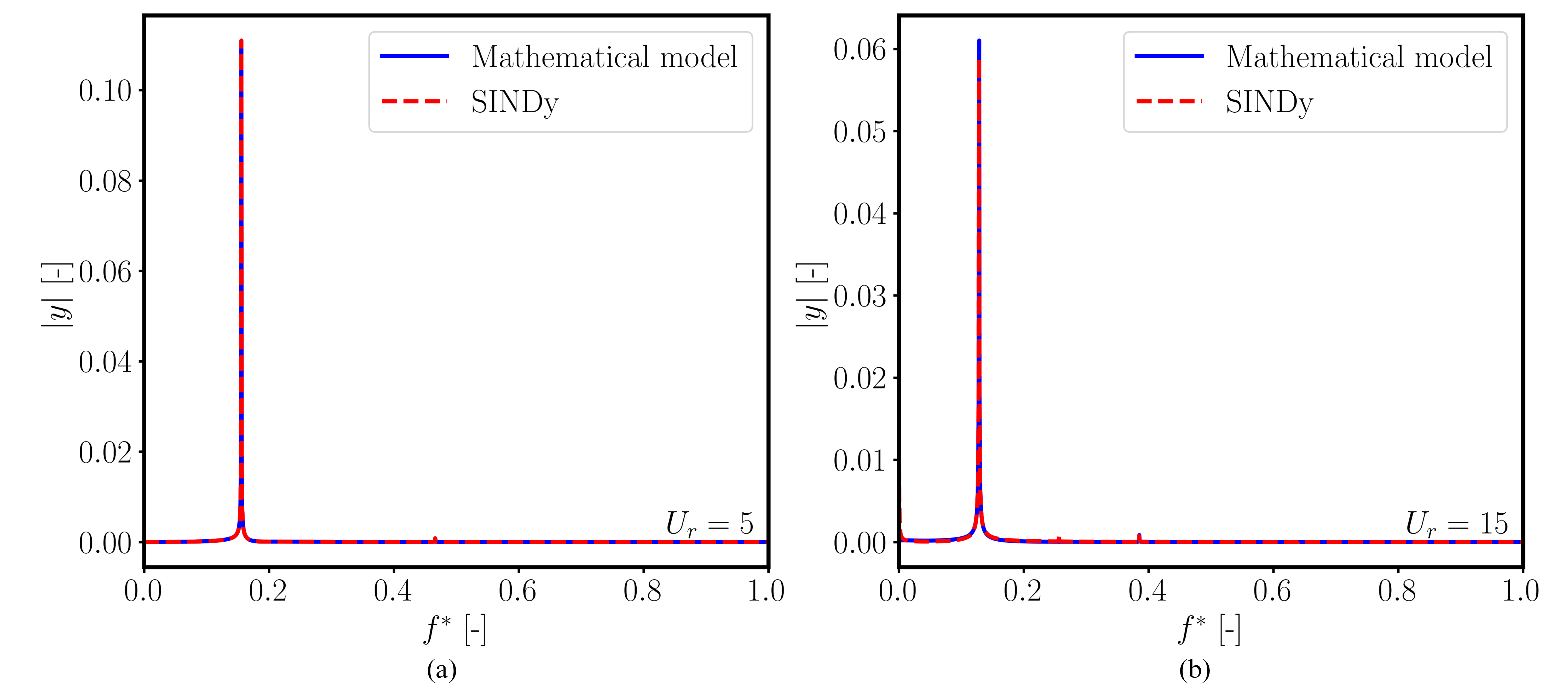}
    \caption{Comparison of dominant oscillation frequencies of $y(=Y/D)$ from SINDy and numerical integration for (a) $U_r = 5$ and (b) $U_r = 15$.}
    \label{fig:fftplot}
\end{figure*}
Figure~\ref{fig:y1q1short} compares the time histories of the structural displacement $y_1$ and wake variable $q_1$ between the numerical and SINDy solutions at $U_r = 5$ and $U_r = 15$. At $U_r = 5$, which lies within the lock-in regime, the agreement is excellent, with the identified model accurately reproducing both the amplitude and phase of oscillations in both state variables. At $U_r = 15$, however, the correspondence deteriorates, particularly in the phase trajectory of $q_1$, suggesting that the identified wake equation deviates from the true dynamics as the system transitions toward the post-lock-in aperiodic regime. This behaviour is consistent with the amplitude error trends observed in Fig.~\ref{fig:errorplot}(b) and may be attributable to the use of a constant sparsity threshold across all reduced velocities. Given that the candidate library includes higher-order terms than required by the synthetic dynamics, variations in coefficient scaling with $U_r$ may influence sparsity selection during model identification, leading either to the retention of superfluous terms or to the pruning of dynamically relevant ones. Notwithstanding these phase discrepancies, Fig.~\ref{fig:fftplot} demonstrates that the dominant oscillation frequency of $y_1$ predicted by SINDy remains in close agreement with the numerical solution at both $U_r = 5$ and $U_r = 15$. This indicates that whilst the identified models exhibit phase drift at higher reduced velocities, they nonetheless preserve the fundamental spectral characteristics of the VIV response, suggesting that the sparse regression procedure successfully captures the dominant energy-containing dynamics even when full trajectory fidelity is not achieved.

In summary, the results of this section demonstrate that standard SINDy can effectively recover the governing dynamics of the coupled 1-DoF wake oscillator system directly from time-series data, without any a priori knowledge of the equation structure, across the majority of the reduced-velocity range examined. The identified models accurately reproduce the structural displacement amplitude and the dominant oscillation frequency throughout the pre-lock-in and lock-in regimes, confirming that sparse regression over a polynomial library is sufficient to capture the essential nonlinear coupling between the structural and wake subsystems under periodic and near-periodic forcing. Performance degradation is observed at higher reduced velocities, particularly for $U_r \geq 15$, and is attributed to the increasing dynamical complexity of the post-lock-in regime, where stronger amplitude modulation and broadband spectral content exceed the representational capacity of a fixed-parameter, identical-polynomial library. This limitation is not intrinsic to the SINDy framework itself but rather reflects the sensitivity of the identification procedure to library construction, suggesting that adaptive basis selection strategies, regime-specific polynomial degrees and sparsity parameters, or the inclusion of physically motivated nonlinear terms could meaningfully improve model fidelity in the aperiodic regime.

These findings establish a rigorous baseline for the full-system identifiability of the coupled VIV problem under idealised conditions and motivate the transition to the more challenging identification task addressed in the following section. Notably, the performance degradation of standard SINDy at higher reduced velocities, where the VIV response undergoes dynamical transitions, highlights a fundamental vulnerability of strong-form sparse 
regression. This sensitivity is expected to be significantly more severe when standard SINDy is applied to high-fidelity CFD data, where the time 
series are additionally subject to nonlinear dynamical transition, providing the primary motivation for the weak-form integral reformulation of WSINDy, which can recover physically consistent governing equations. Sec.~\ref{sec:4} extends the identification campaign to high-fidelity CFD data, where the forcing is broadband, the time series are subject to numerical noise, and no ground-truth governing equations are available for reference, providing a significantly more demanding test of the comparative performance of standard SINDy and WSINDy for the VIV governing equation 
discovery.

\section{VIV system identification from high-fidelity CFD data}
\label{sec:4}
\noindent In this section, high-fidelity CFD simulations are carried out to generate training data for the SINDy and WSINDy identification frameworks, providing a physically consistent dataset that complements the synthetic wake-oscillator data examined in the preceding section. Unlike the low-order wake oscillator model, which relies on empirically calibrated parameters and a prescribed equation structure, the CFD simulations resolve the full Navier-Stokes dynamics of the coupled fluid-structure system without modelling assumptions on the wake behaviour, thereby producing time-series data that captures the broadband spectral content, nonlinear amplitude modulation, and regime-dependent dynamics characteristic of real VIV responses. 
\begin{figure}[htbp]
    \centering
    \includegraphics[width=0.72\textwidth]{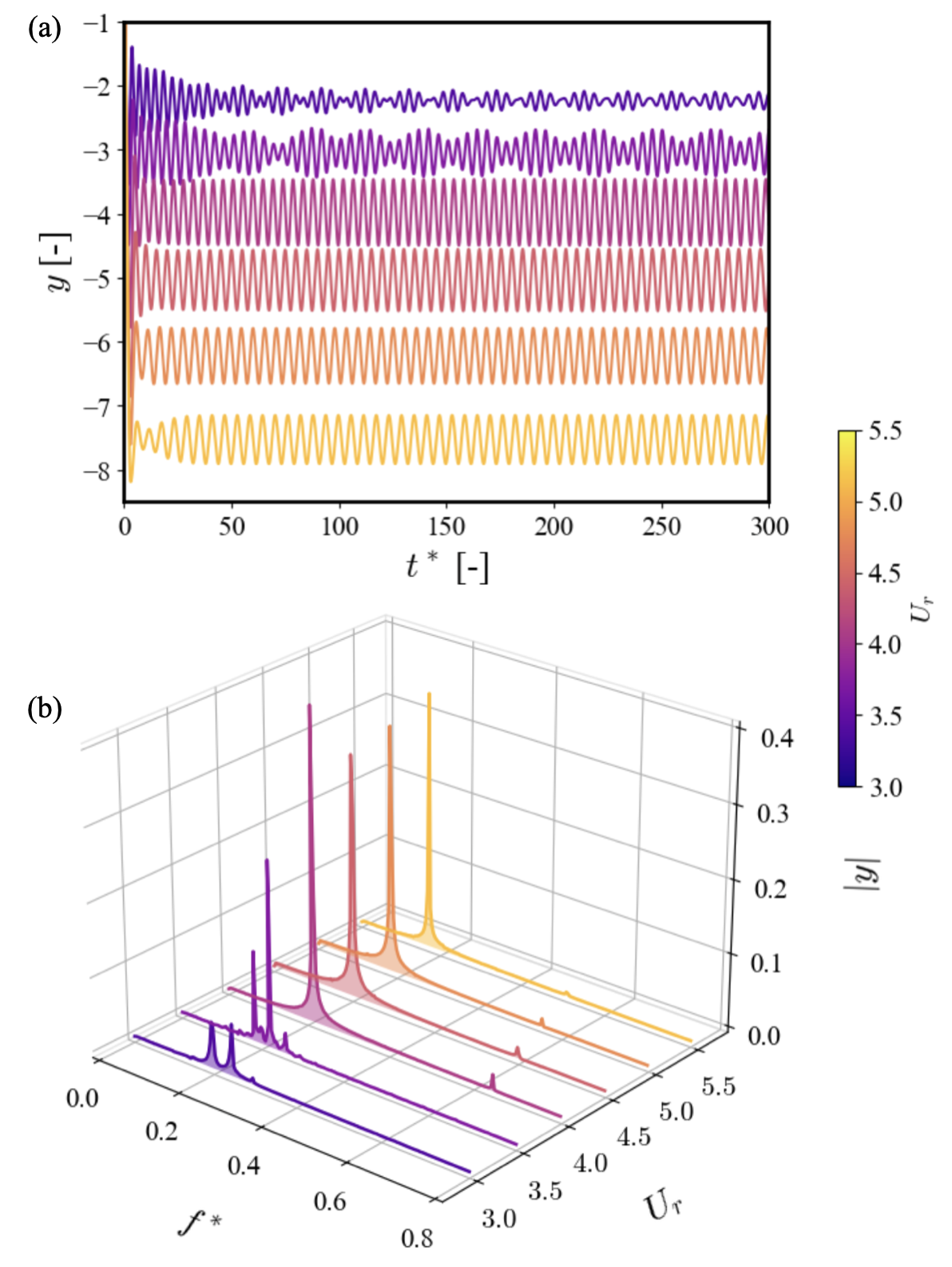}
    \caption{CFD results for the transverse displacement of the elastically mounted 1-DoF circular cylinder across reduced velocities $U_r \in [3.0, 5.5]$. (a) Time histories of the dimensionless transverse displacement $y$ as a function of dimensionless time $t^*$, coloured by reduced velocity $U_r$. (b) Three-dimensional representation of the fast Fourier transform (FFT) spectral content of $y$, showing the amplitude spectrum $|y|$ as a function of dimensionless frequency $f^*$ for each $U_r$, illustrating the evolution of the dominant shedding frequency across the reduced-velocity range.}
    \label{fig:CFD_timeseries_FFT}
\end{figure}
The simulations concern the transverse VIV of an elastically mounted 1-DoF circular cylinder across a range of reduced velocities, encompassing the pre-lock-in, lock-in, and post-lock-in regimes, and the resulting force and displacement time series are subsequently used as input to the sparse identification algorithm. The detailed description of the computational methodology and the verification and validation of the computational setup are presented in \ref{app:B}.

Figure~\ref{fig:CFD_timeseries_FFT} presents the transverse displacement response of the elastically mounted cylinder, obtained from the CFD simulations, across reduced velocities $U_r \in [3.0, 5.5]$. Figure~\ref{fig:CFD_timeseries_FFT}(a) shows the time histories of the dimensionless displacement $y$, from which it is evident that the response attains a statistically stationary limit-cycle oscillation within the early stages of the simulation ($t \geq 50$) across all reduced velocities considered, with the oscillation amplitude increasing monotonically with $U_r$. The time series exhibits periodicity at $U_r \geq 4$, indicative of the onset of the lock-in regime. Figure~\ref{fig:CFD_timeseries_FFT}(b) presents the corresponding FFT amplitude spectra of $y$ for each $U_r$, revealing a dominant single-frequency response at $U_r \geq 4$. The dominant dimensionless shedding frequency $f^* (=\frac{f D}{U_\infty})$ is observed to decrease with increasing $U_r$ as the vortex shedding frequency approaches and synchronises with the structural natural frequency during lock-in. An exception to the single-frequency behaviour is observed at the lowest reduced velocities considered, $U_r = 3.0$ and $U_r = 3.5$, which lie within the pre-lock-in regime. At these reduced velocities, the FFT spectra exhibit multiple distinct peaks rather than a single dominant frequency, indicating quasi-periodic oscillations arising from the competition between the vortex-shedding frequency and the structural natural frequency prior to synchronisation. This multi-frequency spectral content reflects the absence of full frequency lock-in and poses a significantly more demanding identification challenge for sparse regression, thereby increasing the susceptibility of standard SINDy to derivative-estimation error and further motivating the application of WSINDy in this regime.

\begin{figure*}[htbp]
    \centering
    \includegraphics[width=1\textwidth]{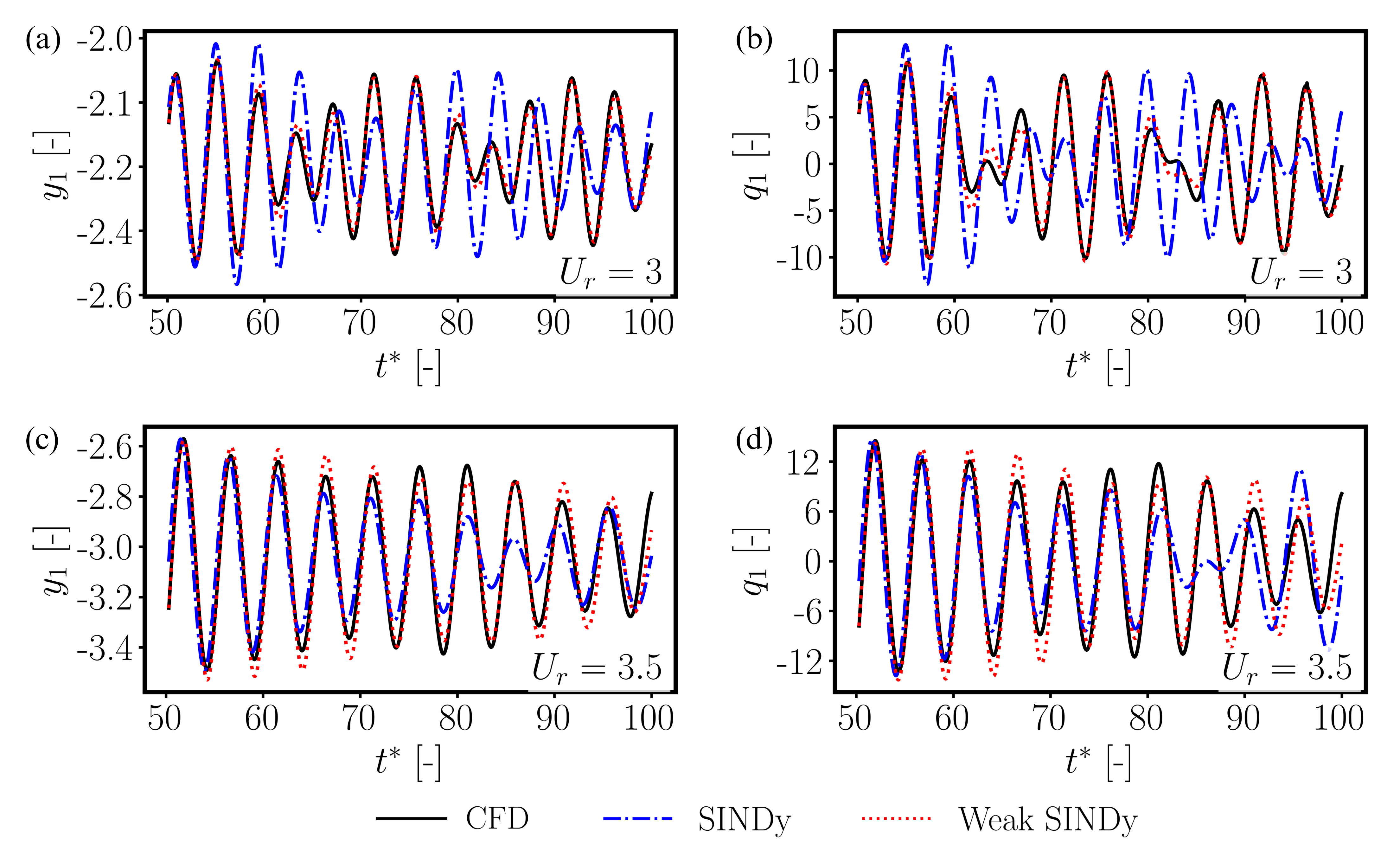}
   \caption{Comparison of the CFD simulation, SINDy, and WSINDy reconstructions for the 1DOF single-cylinder vortex-induced vibration (VIV) system in aperiodic regime. The time histories of the transverse displacement $y_1$ and the wake variable $q_1$ are shown for (a–b) $U_r = 3$, (c–d) $U_r = 3.5$.}
 \label{fig:traj1dofaperiod}
\end{figure*}

We first analyse the pre-lock-in aperiodic regime, concentrating on reduced velocity values, \( U_r \in \{3,\, 3.5\} \). SINDy and WSINDy are used to identify the governing low-order model from CFD-generated temporal dynamics. The temporal variables are chosen based on the reduced model discussed in Sec.~\ref{app:A}, where the entire system is governed by the coupled dynamics of \( y_1 \), \( y_2 \), \( q_1 \), and \( q_2 \). Here, \( y_1 \) and \( y_2 \) are the structural displacement and its derivative, and \( q_1 \) and \( q_2 \) are the wake parameter and its derivative. The wake variable is indirectly obtained from the lift force \( C_L \) by using the transformation:
\begin{equation}
q_1 = 2C_L/C_{L0},
\end{equation}
where \(C_{L0}\) is 0.3 in the large range of Reynolds number \( (300 < \text{Re} < 1.5 \times 10^5) \) as reported by \citet{Facchinetti}. This step is crucial because it enables a data-driven reconstruction of the wake-oscillator model. However, the direct \( C_L \) signal obtained from the CFD data consists of high-frequency fluctuations due to numerical noise. To mitigate this and obtain a smoother representation of the wake variable, a moving average filter is applied to the \( C_L \) time series before further processing. Figure~\ref{fig:traj1dofaperiod} presents a comparison of the time histories of the structural displacement $y_1$ and wake variable $q_1$ reconstructed by standard SINDy and WSINDy against the CFD reference data at $U_r = 3.0$ and $U_r = 3.5$, both of which lie within the pre-lock-in aperiodic regime characterised by quasi-periodic oscillations and multi-frequency spectral content. Both identification methods use a second-degree polynomial library, and the results reveal a marked difference in reconstruction fidelity between the two formulations, which is directly attributable to their contrasting approaches to derivative estimation.

Standard SINDy exhibits substantial deviations from the CFD reference in both $y_1$ and $q_1$ at both reduced velocities, with the predicted trajectories accumulating phase errors and amplitude discrepancies over the observation window despite reproducing the qualitative character of the oscillations. This deterioration is consistent with the derivative-estimation sensitivity of strong-form sparse regression identified in Sec.~\ref{sec:2.1}: in the pre-lock-in regime, the multi-frequency nature of the state trajectories renders pointwise numerical differentiation particularly unreliable, and the resulting errors in $\dot{X}$ propagate directly into the identified coefficient matrix $\Xi$, producing a model that drifts from the true dynamics under forward integration. A third-degree polynomial library was also evaluated, but produced unstable identified models in both methods, with noise amplification in higher-order cross-interaction terms causing divergence under simulation, confirming that the library complexity must be balanced against the noise level inherent in the data.

WSINDy, by contrast, demonstrates markedly superior reconstruction fidelity at both $U_r = 3.0$ and $U_r = 3.5$, with the predicted time histories of $y_1$ closely tracking the CFD reference in both amplitude and phase throughout the observation window. The improvement is less pronounced for the wake variable $q_1$, for which both methods exhibit greater discrepancy relative to the CFD reference. This disparity arises because the wake dynamics are derived indirectly from the lift coefficient $C_L$, which, despite preprocessing with a moving average filter, retains residual high-frequency noise. Nonetheless, WSINDy produces a measurably more faithful reconstruction of $q_1$ than standard SINDy, confirming that the weak-form reformulation provides a meaningful and consistent improvement in identification accuracy across both state variables in the aperiodic pre-lock-in regime.

\begin{figure*}[htbp]
    \centering
    \includegraphics[width=1\textwidth]{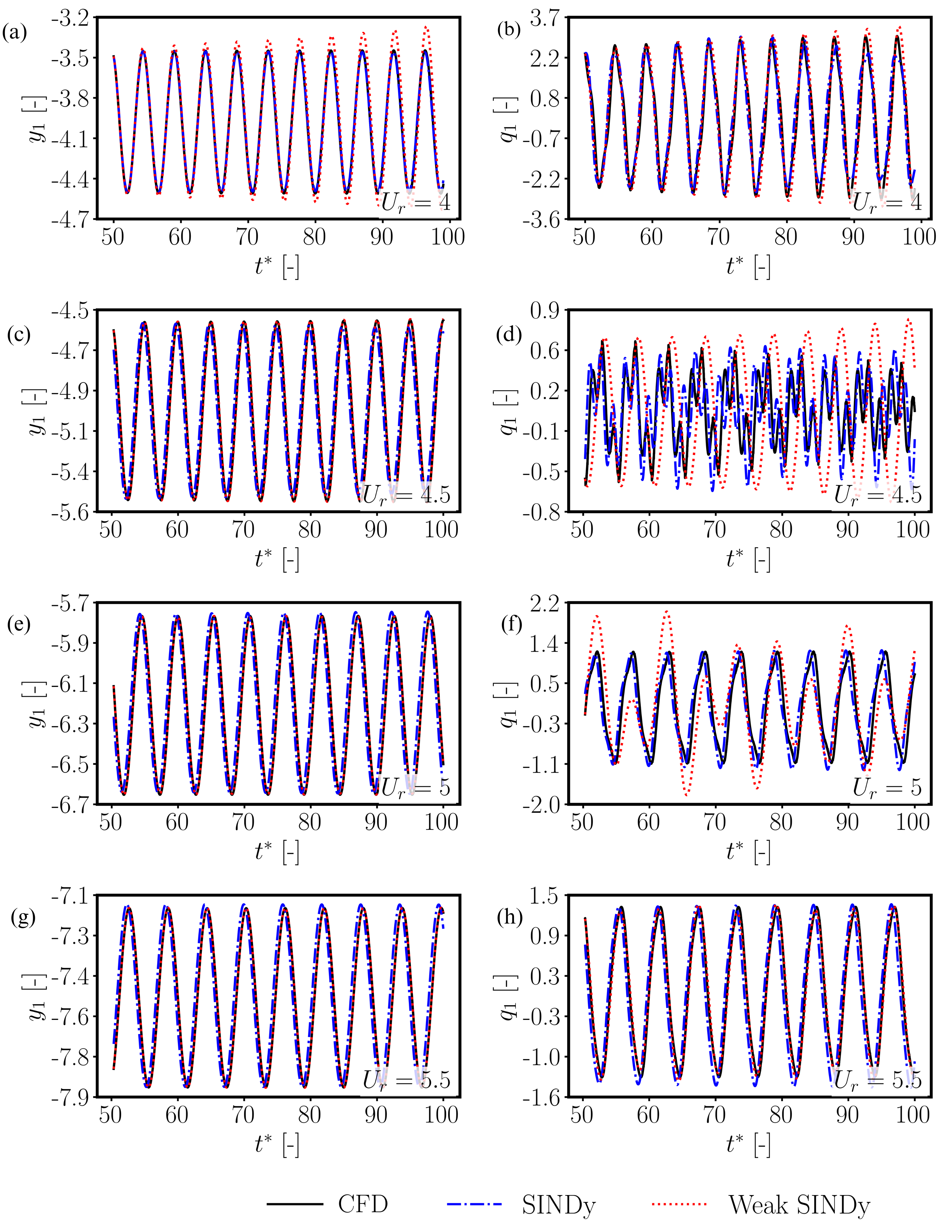}
    \caption{Comparison of the CFD simulation, SINDy, and WSINDy reconstructions for the 1-DOF single-cylinder vortex-induced vibration (VIV) system in periodic regime. The time histories of the transverse displacement $y_1$ and the wake variable $q_1$ are shown for (a–b) $U_r = 4$, (c–d) $U_r = 4.5$, (e–f) $U_r = 5$, (g–h) $U_r = 5.5$.}
    \label{fig:traj1dofperiod}
\end{figure*}

Next, we analyse the performance of both the SINDy algorithms for the lock-in regime corresponding to \( U_r \in \{4,\, 4.5,\, 5,\, 5.5\} \), where vortex shedding is synchronised with structural motion, and the 
response is characterised by large-amplitude, near-sinusoidal 
oscillations; see Fig.~\ref{fig:traj1dofperiod}. In the structural displacement $y_1$, both SINDy and WSINDy achieve excellent agreement with the CFD reference across all four reduced velocities, accurately reproducing the amplitude, phase, and waveform of the periodic response with a second-degree polynomial library. This confirms that the regular SINDy model is sufficient to capture the single-frequency, limit-cycle response. At $U_r = 4.0$, both methods recover $q_1$ accurately, but at $U_r = 4.5$, the wake variable exhibits irregular, low-amplitude oscillations in the CFD data that neither method fully reproduces, with standard SINDy showing larger deviations than WSINDy. At $U_r = 5.0$ and $U_r = 5.5$, the agreement in $q_1$ recovers for both methods, with WSINDy providing marginally superior phase fidelity.

When the polynomial library degree was increased to explore the recovery of higher-order nonlinear terms, the identified models produced by both methods exhibited instability under forward simulation, with higher-order cross-interaction terms amplifying small integration errors and causing divergence over long time horizons. This instability was not observed when the same library was applied to the noise-free synthetic wake-oscillator data in Sec.~\ref{sec:3}, confirming that even low-level numerical noise present in CFD-derived time series can corrupt the coefficient estimates of higher-order library terms, rendering the identified model ill-conditioned. This finding underscores a fundamental trade-off in sparse identification from CFD data: whilst richer libraries offer greater representational flexibility, they simultaneously increase sensitivity to noise in both derivative estimates and feature evaluations, necessitating either explicit regularisation or prior physical knowledge to constrain the active term set.

\begin{figure*}[htbp]
    \centering
    \includegraphics[width=1\textwidth]{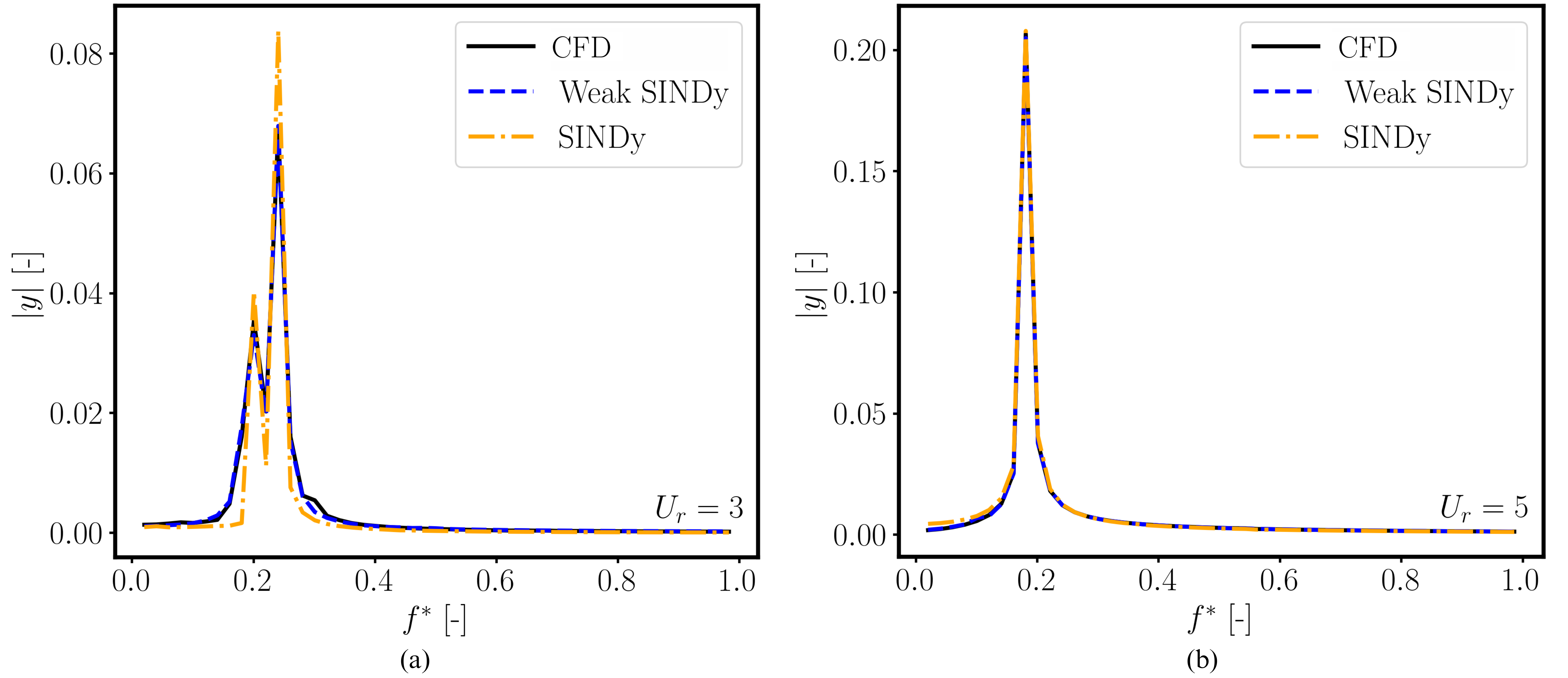}
    \caption{Dominant oscillation frequencies of $y$ predicted by SINDy and WSINDy in comparison with CFD results for (a) $U_r = 3$ and (b) $U_r = 5$.}
   \label{fig:fftplotCFD}
\end{figure*}

Figure~\ref{fig:fftplotCFD} presents the FFT amplitude spectra of the structural displacement $y$ predicted by SINDy and WSINDy in comparison with the CFD reference at $U_r = 3.0$ and $U_r = 5.0$. At $U_r = 5.0$, both methods reproduce the dominant shedding frequency with high accuracy, with the predicted spectral peaks matching the CFD reference precisely, confirming that the identified models accurately capture the single-frequency limit-cycle dynamics of the lock-in regime. At $U_r = 3.0$, the CFD spectrum exhibits the multiple closely spaced peaks characteristic of quasi-periodic pre-lock-in oscillations. Both methods reproduce the approximate location of the dominant-frequency cluster, but WSINDy yields a cleaner spectral reconstruction, with peak locations and relative amplitudes that more closely match the CFD data. This result provides further quantitative evidence that WSINDy offers a meaningful and consistent improvement over standard SINDy in the aperiodic pre-lock-in regime.

\begin{figure}[htbp]
    \centering
    \includegraphics[width=0.8\linewidth]{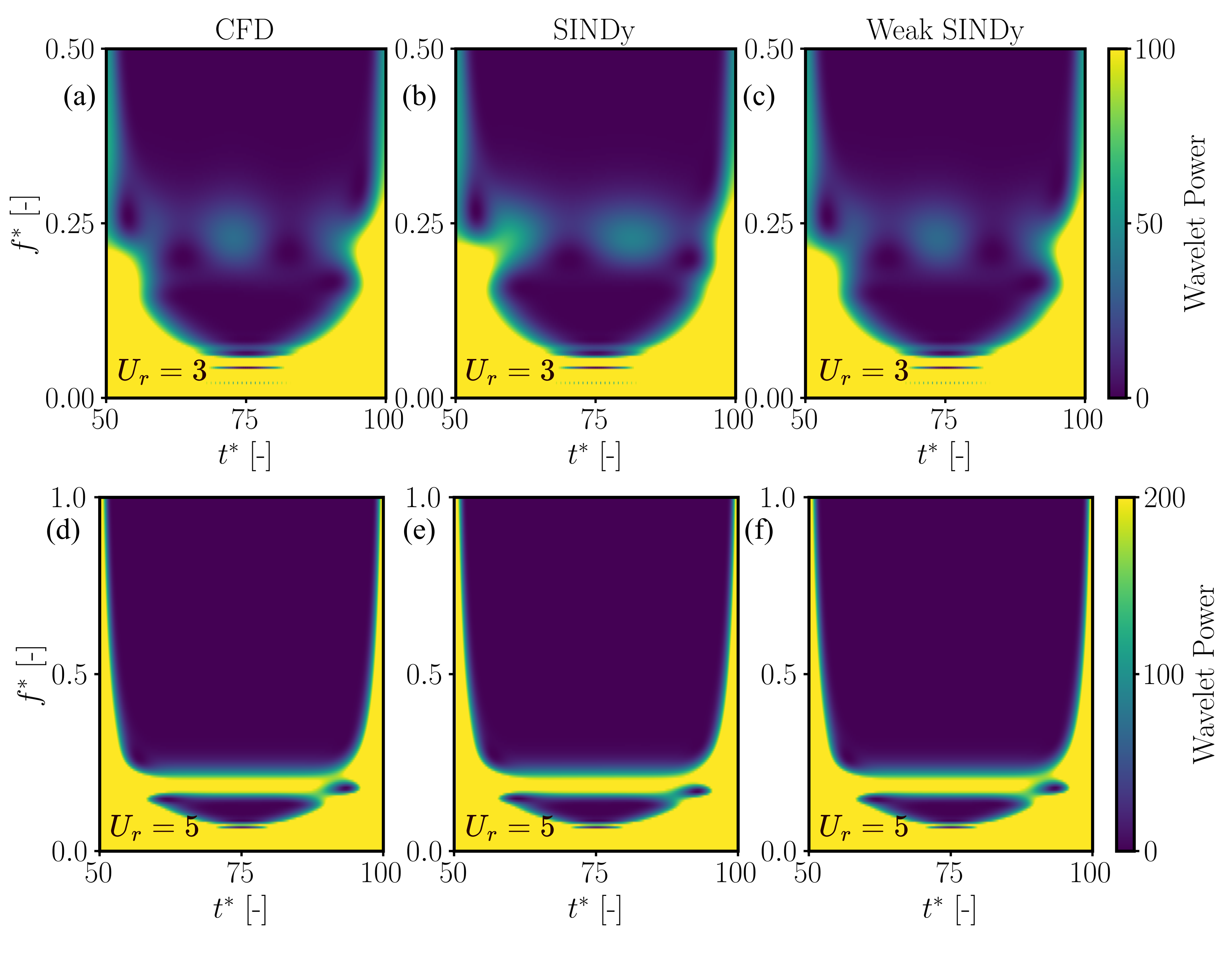}
    \caption{Comparison of continuous Morlet wavelet spectra for the VIV response, obtained from CFD, SINDy, and Weak-SINDy models. Panels (a)–(c) correspond to $U_r = 3$, and panels (d)–(f) correspond to $U_r = 5$. The colour scale indicates wavelet power.}
    \label{fig:wavelet_comparison}
\end{figure}

\begin{figure}[htbp]
    \centering
   \includegraphics[width=0.6\textwidth]{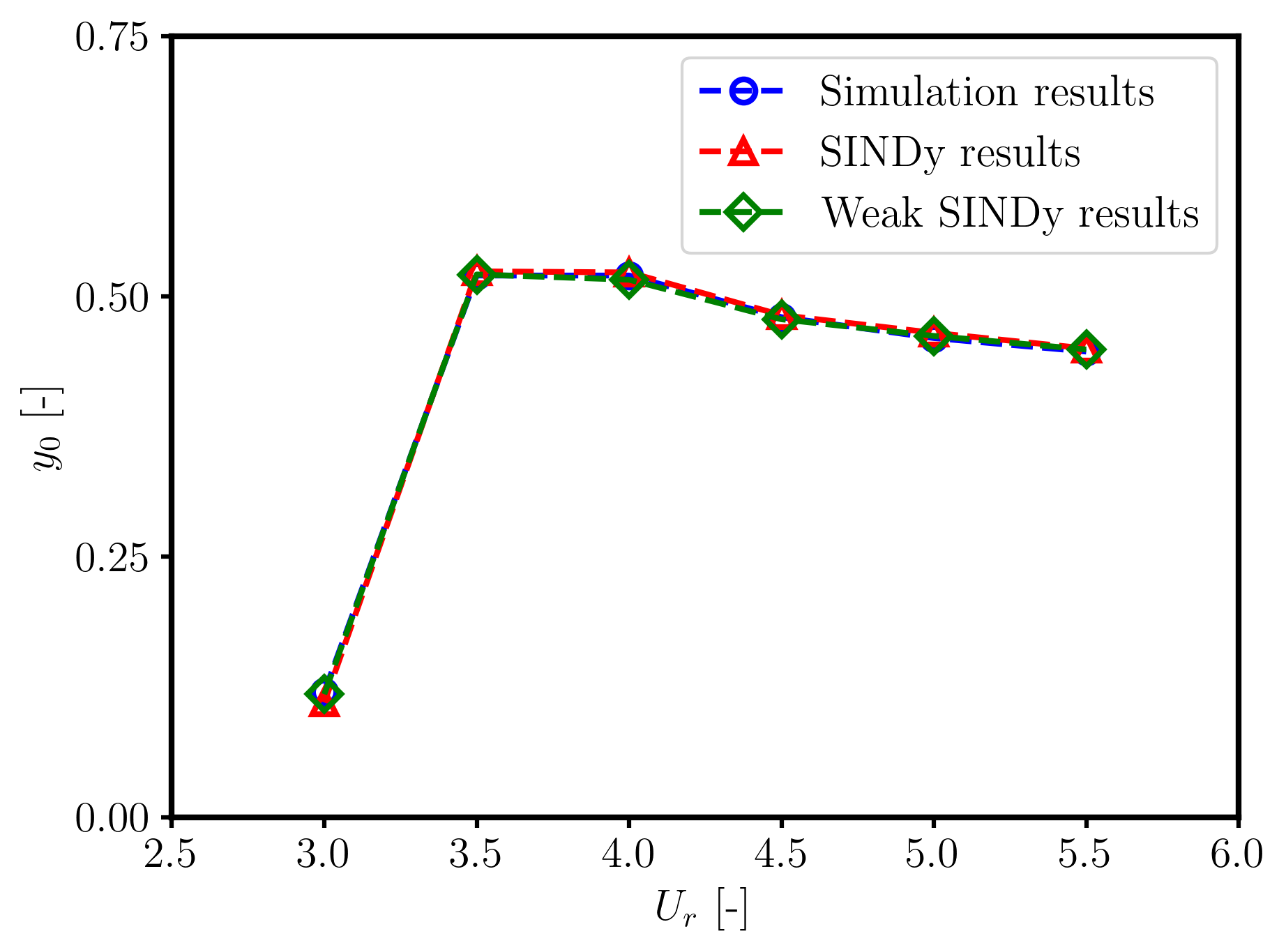}
    \caption{Comparison of vortex-induced vibration (VIV) amplitude response obtained from the present CFD-based results, SINDy, and WSINDy models across $U_r$.}
    \label{fig:amplituderecons}
\end{figure}

Further, Fig. \ref{fig:wavelet_comparison} presents the continuous Morlet wavelet spectra of the transverse displacement $y_1(t)$ obtained from CFD, SINDy, and Weak-SINDy at $U_r = 3$ and $U_r = 5$. The case $U_r = 3$ corresponds to a relatively aperiodic regime, where the dominant frequency is located slightly below $f = 0.25$ and exhibits mild temporal variations. Although SINDy captures the overall oscillatory nature of the response, small discrepancies are observed in the distribution and intensity of wavelet power near the dominant ridge when compared to CFD. WSINDy provides improved agreement, reproducing both the frequency localisation and temporal evolution of the dominant mode more accurately. In contrast, $U_r = 5$ represents a strongly periodic regime characterised by a stable oscillation with the dominant frequency centred at $f = 0.2$. In this case, both SINDy and Weak-SINDy closely match the CFD spectrum, accurately capturing the sharp and persistent spectral ridge. These findings are consistent with the FFT results shown in Fig. ~\ref{fig:fftplotCFD}, where the dominant frequency concentrates near $0.25$ for $U_r = 3$ and exactly at $0.2$ for $U_r = 5$, confirming agreement between the time-frequency and frequency-domain analyses across both periodic and aperiodic regimes. Fig. \ref{fig:amplituderecons} compares the VIV amplitude response from the present CFD-based results with SINDy and WSINDy models across reduced velocities $U_r$. The amplitude predictions from both data-driven models appear highly accurate, closely matching the CFD results. However, a close observation of Figs.~\ref{fig:traj1dofaperiod} and \ref{fig:traj1dofperiod} reveals minor phase shifts and predicted trajectory deviations. To provide better insight, the root-mean-square errors (RMSE) of the predicted results are discussed next.

\begin{figure*}[htbp]
    \centering
    \includegraphics[width=0.6\textwidth]{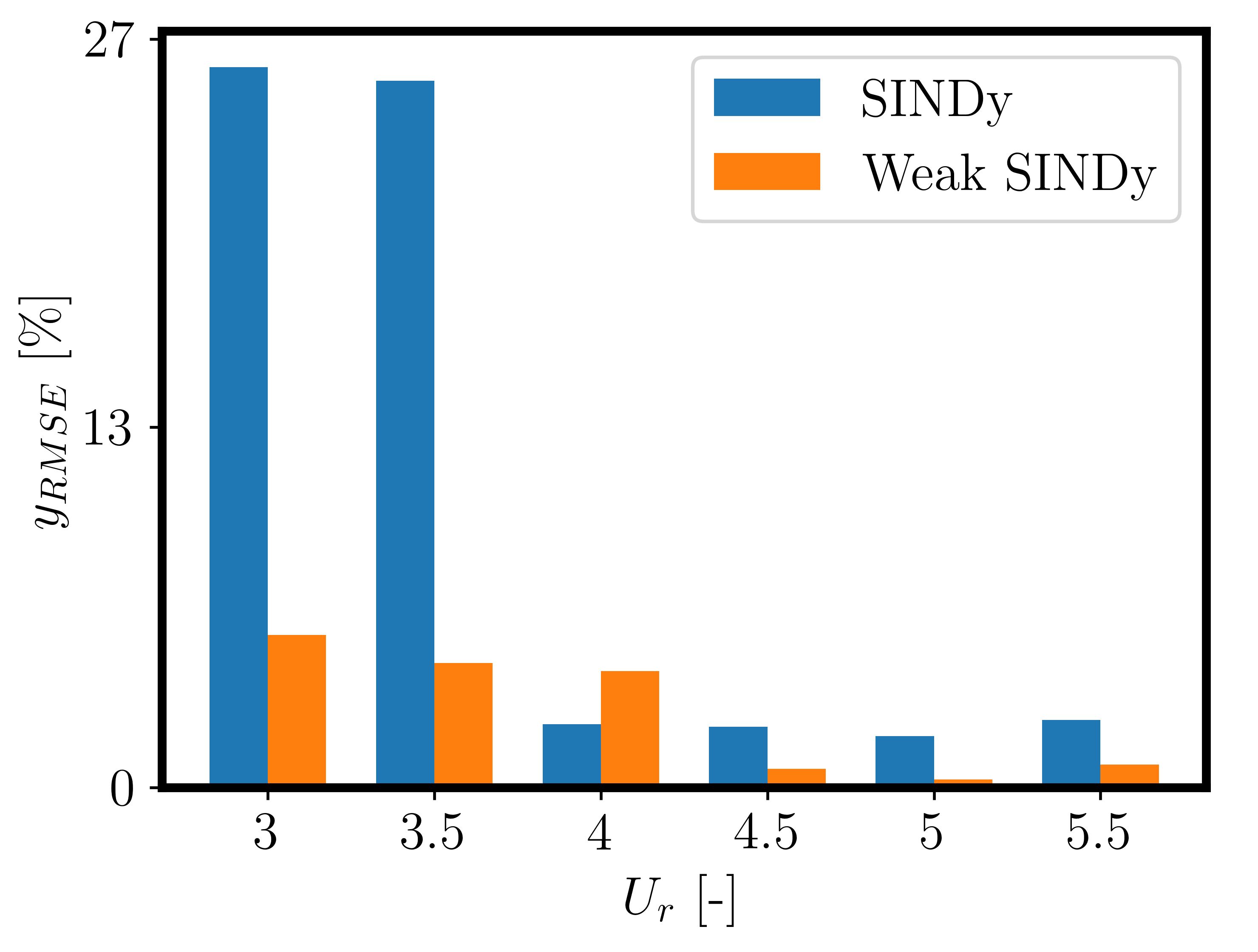}
    \caption{Comparison of RMSE for SINDy and WSINDy across different reduced velocities \(U_r\). Results are shown for non-dimensional displacement \(y\). The analysis was performed on time-series data with a time window of 50 (from $t^* = 50$ to $100$).}
\label{fig:RMSEbar_single}
\end{figure*}

\begin{figure*}[htbp]
    \centering
    \includegraphics[width=0.8\textwidth]{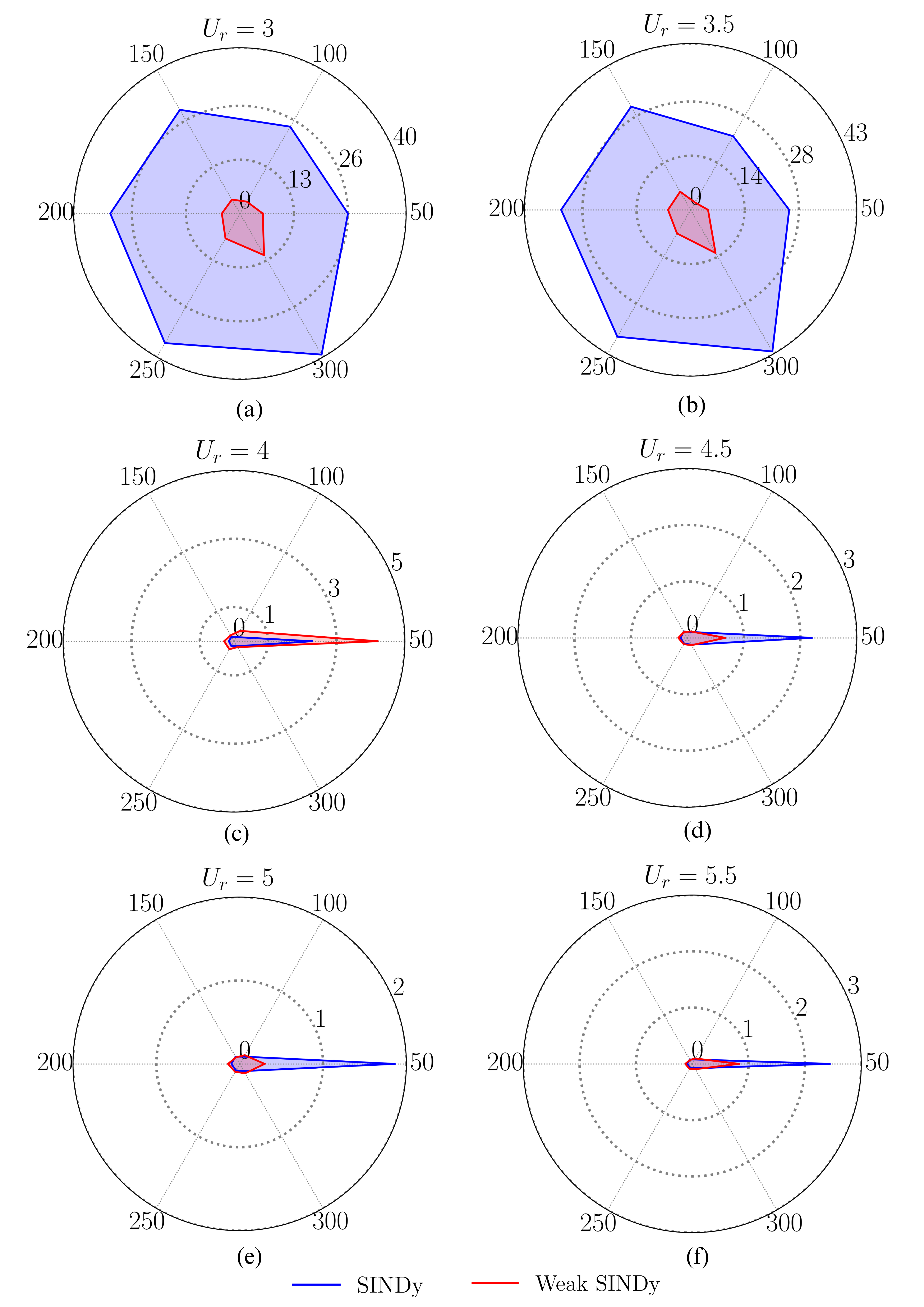}
    \caption{Comparison of relative errors in structural displacement response between SINDy (blue) and WSINDy (red) models at different reduced velocities \(U_r\). 
    Each radar plot corresponds to a specific \(U_r\) and shows the RMSE expressed as a percentage, computed over increasing time windows starting from $t^*=50$. The labels 50, 100, 150, 200, 250, and 300 denote RMSE evaluated over the intervals 
$t^* = 50\text{--}100$, $50\text{--}150$, $50\text{--}200$, $50\text{--}250$, $50\text{--}300$, and $50\text{--}350$, respectively. 
    (a) \(U_r = 3\); (b) \(U_r = 3.5\); (c) \(U_r = 4\); (d) \(U_r = 4.5\); (e) \(U_r = 5\); (f) \(U_r = 5.5\).}
    \label{fig:spydersingle}
\end{figure*}

\begin{figure*}[htbp]
    \centering
    \includegraphics[width=1.1\textwidth]{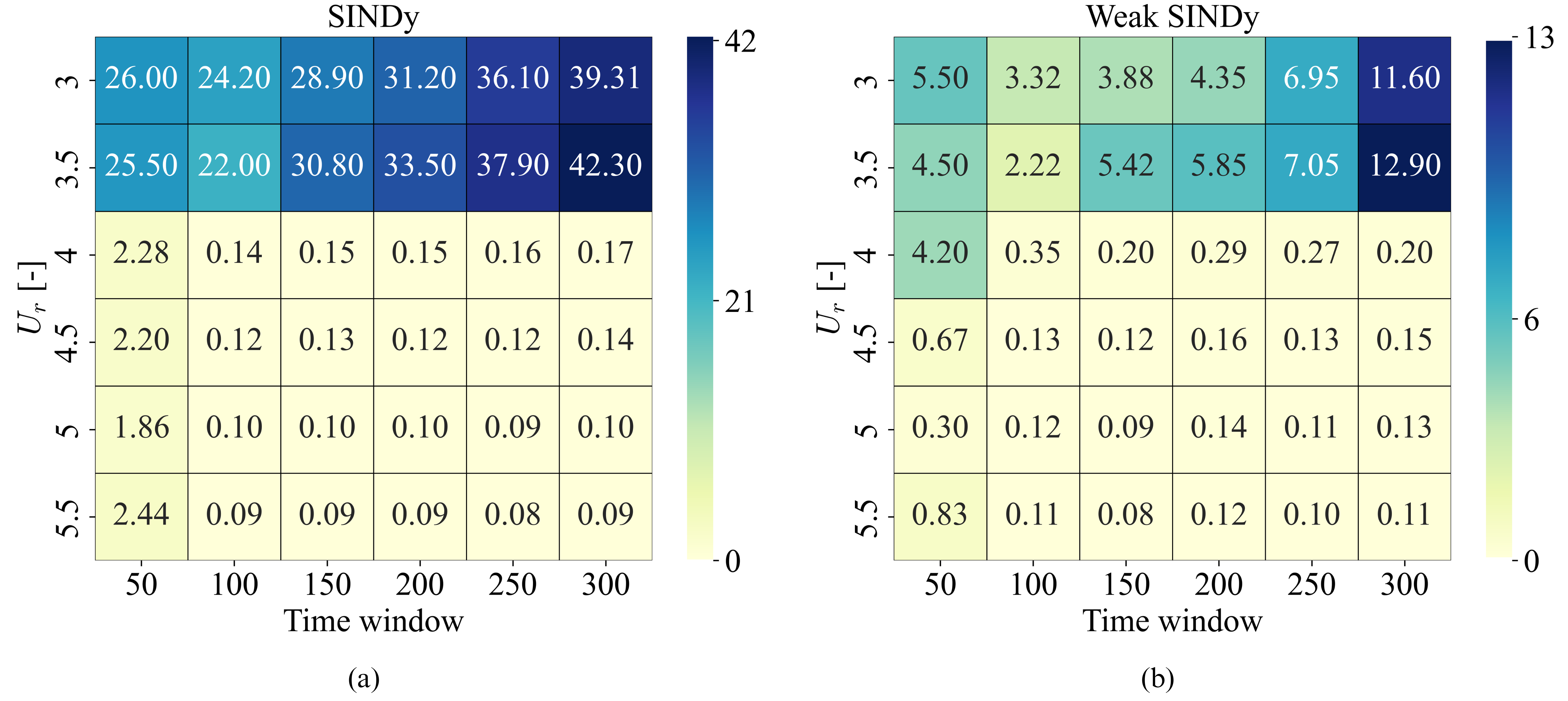}
    \caption{Comparison of prediction errors (colour-coded) for different time windows starting from $ t^*$ = 50 and reduced velocities $U_r$ using (a) SINDy and (b) WSINDy models.}
    \label{fig:heatmap_single}
\end{figure*}

Figures~\ref{fig:RMSEbar_single}--\ref{fig:heatmap_single} summarise the performance of SINDy and WSINDy models across different time series lengths and reduced velocities in terms of the root-mean-square error expressed as a percentage of the mean value. In this study, greater emphasis is placed on predicting structural displacement, as it directly characterises the cylinder response due to wake-induced vibrations. Hence, we compare how increasing data will affect the prediction of $y$ with minimum data availability. Figure~\ref{fig:RMSEbar_single} shows the error for structural displacement $y$ across different reduced velocities using both SINDy and WSINDy. This plot shows the error for an analysis based on a limited dataset of 5,000 points, spanning the non-dimensional time window from $t^* (= \frac{tU_\infty}{D}) = 50$ to $100$ at a record interval of 0.01, highlighting model performance under limited data availability. At $t^* < 50$, the wake is in a transient state as the flow adjusts, and vortex shedding has not yet been fully established. This portion of the data reflects the initial development of the wake and is therefore excluded from the study. For the structural displacement, WSINDy clearly outperforms SINDy in the aperiodic regime at $U_r=3$ and $U_r=3.5$, reducing errors by approximately 20\%. This underscores the importance of the weak form for identifying irregular dynamics. Beyond this regime, the two methods perform similarly, with WSINDy having a slight edge due to better phase identification.

Figure~\ref{fig:spydersingle} illustrates how the error evolves with increasing data length, where the length of the time window is counted starting from 50 recorded at a 0.01 interval, which corresponds to the data from $t^* = 50$ to $100$. For the time window of 100, we consider the data from $t^* = 50$ to $150$, and so on. The radial distance in each polar plot corresponds to the percentage root-mean-square error. In the aperiodic regime ($U_r = 3$ and $3.5$), the error for the regular SINDy-predicted response initially drops slightly from time window 50 to 100, but then increases rapidly as more data are added, reaching nearly 50\% for longer datasets. This shows that SINDy struggles to handle irregular, unpredictable dynamics in which multiple frequencies dominate. WSINDy exhibits a similar trend, but the errors are significantly lower and remain stable, indicating that it’s much better at capturing aperiodic behaviour. Increasing the length of the time histories increases data irregularity, and the algorithm struggles to identify dynamics in which multiple frequencies co-dominate. It is crucial to note here that SINDy can often tend to overfit on irregularities rather than the evolution of the trajectory due to its derivative estimation method, making identification less physically relevant, and error tends to creep up when simulating the model for longer time windows, some of which can be an effect of numerical instability. In the periodic regime ($U_r =$ 4 to 5.5), both methods start with slightly higher errors at 50 steps, improve over the next 100 steps, and then errors creep back up from 150 steps onward. WSINDy consistently remains below SINDy, but the overall pattern indicates that beyond a certain data length, additional data do not necessarily improve performance. Overall, WSINDy is more robust to irregular dynamics, even as identifying dynamics becomes challenging with increasing data. Figure~\ref{fig:heatmap_single} provides a colour-coded summary of the error of SINDy and WSINDy for different reduced velocities and time series lengths, giving an overall summary of model performance trends.

\begin{figure}[htbp]
    \centering
    \includegraphics[width=\textwidth]{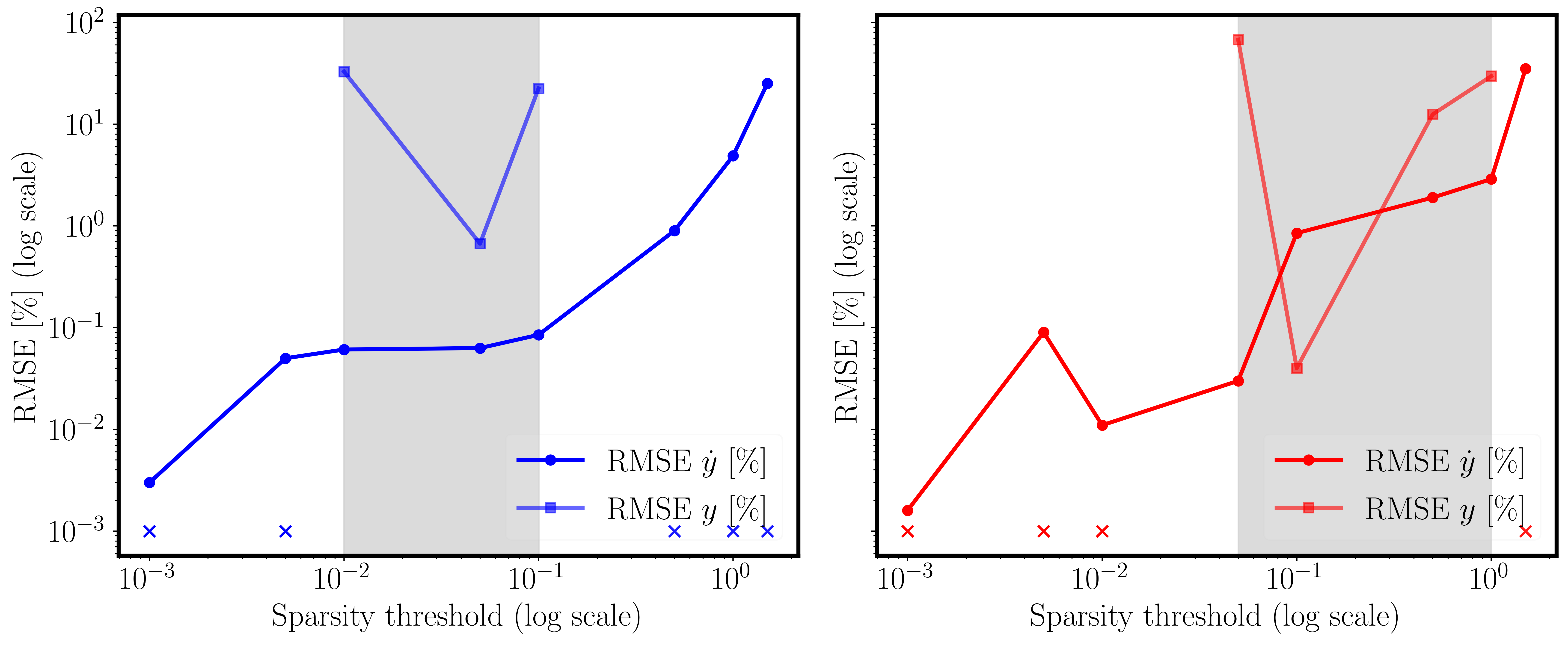}
    \caption{RMSE comparison for $y$ and $\dot{y}$ using SINDy (left) and WSINDy (right) models across different sparsity thresholds for $U_r=5$. Both axes are in logarithmic scale.}
    \label{fig:sindy_weak_loglogsparsity}
\end{figure}

\begin{figure}[htbp]
    \centering
\includegraphics[width=0.5\textwidth]{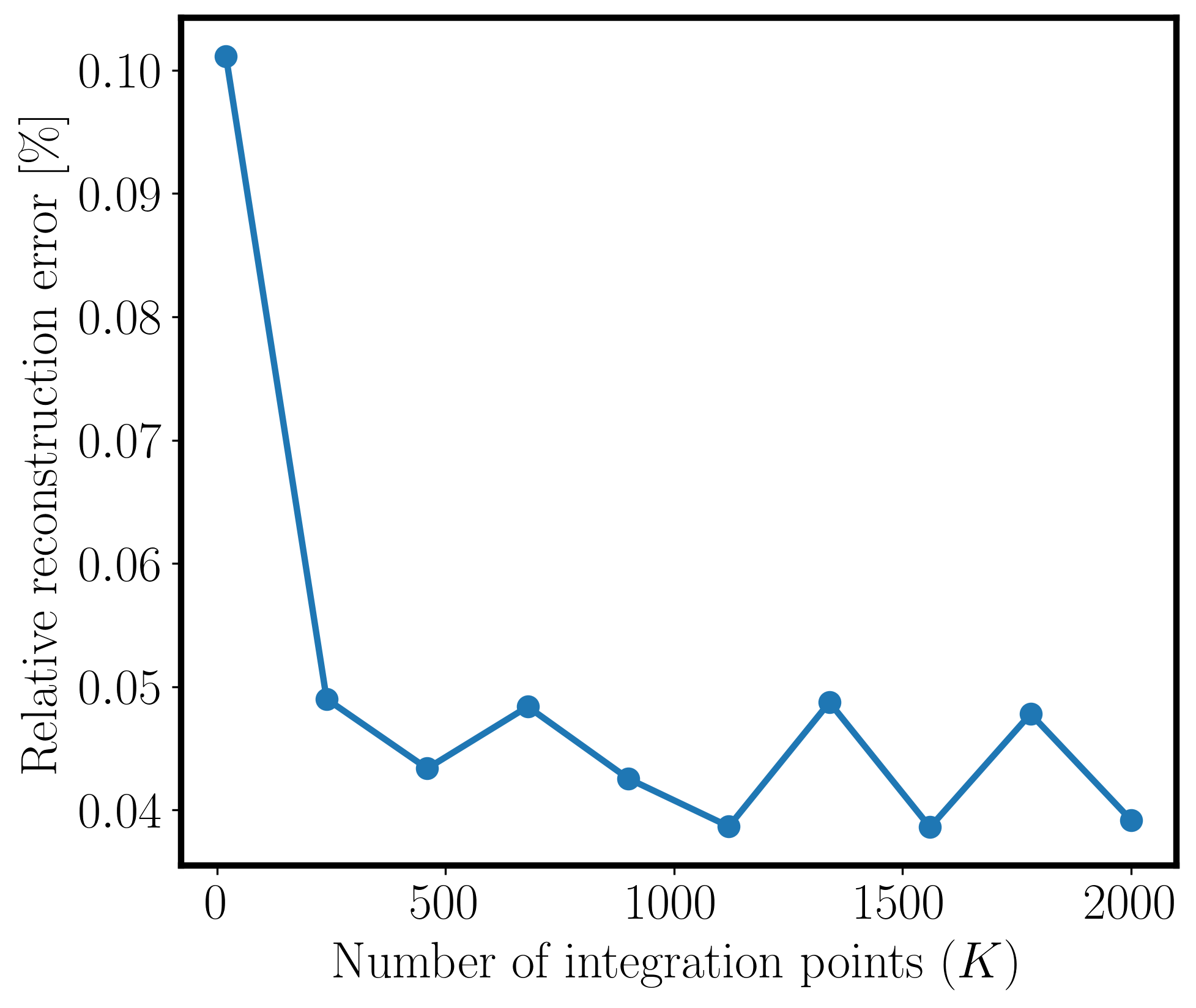}
    \caption{Variation of the relative error $\varepsilon(\%)$ with respect to the number of integration points $K$ used in the WSINDy formulation, illustrating the effect of the numerical integration resolution on model accuracy.}
    \label{fig:integration_error}
\end{figure}

Fig. \ref{fig:sindy_weak_loglogsparsity} illustrates the variation of root mean squared errors with respect to varying sparsity thresholds ($\lambda$) for both the standard SINDy algorithm, implemented with Sequentially Thresholded Least Squares (STLSQ), and WSINDy, implemented using the SR3 optimisation approach. This figure corresponds to results obtained by fitting models to the data for $U_r=5$. 
The only input parameter required for using SINDy with the STLSQ optimiser is $\lambda$. In WSINDy, the relaxation parameter ($w$) is taken as $10^{-2}$ for the SR3 optimiser, and the number of test functions ($K$) is taken as 1500 due to having a minimum relative reconstruction error as shown in Fig. \ref{fig:integration_error}. Therefore, in the weak formulation, the governing equations are projected onto 1500 temporal test functions, or equivalently, subdomains \citep{weakSINDy}.  Each test function defines a localised time window, and integrating the dynamics against all $K$ test functions produces 1500 weak integral constraints that are collectively used to identify the sparse coefficients of the model. Together, the three parameters $\lambda, w, k$ influence how the WSINDy algorithm balances fidelity, sparsity, and robustness against measurement noise.

In this analysis, the sparsity threshold was varied over eight values: $\lambda \in \{0.001, 0.005, 0.01, \\ 0.05, 0.1, 0.5, 1, 1.5 \}$. For SINDy, $\dot{y}$ RMSE percentage varied from $0.003$ to $25.2$, while $y$ RMSE percentage spanned $0.67$ to $32.9$; for WSINDy, $\dot{y}$ RMSE varied from $0.0016$ to $35.2$, while $y$ RMSE spanned $0.04$ to $67.8$. The sparsity thresholds at which the identified models diverged during simulation have been marked with a cross. For example, SINDy becomes unstable at thresholds $0.001$, $0.005$, $0.5$, $1$, and $1.5$, whereas WSINDy remains stable across a broader range, only showing instability at thresholds $0.001$, $0.005$, and $1.5$. The grey-shaded region in the figure highlights the models' stability thresholds. SINDy performed best at $\lambda=0.05$ whereas WSINDy did best at $\lambda=0.1$. The corresponding equations have been shown in Sec. \ref{equations_Ur5}. At extremely small thresholds, the model hardly zeroes out any coefficients, leading to overfitting and a low RMSE for $\dot{y}$; however, the governing equations are extremely unstable due to unnecessary variables. At extremely high thresholds, the model's aggressive pruning of relevant terms results in lower accuracy and model divergence. A possible explanation for the different threshold ranges is that, in standard SINDy, the sparsity threshold must remain relatively low because aggressive pruning can remove dynamically significant terms when using sequential thresholding. However, in WSINDy, the SR3 relaxation term ensures that the sparsity constraint does not over-penalise coefficients, allowing the model to retain essential coefficients even at higher thresholds, enabling stable and physically meaningful dynamics over a broader sparsity range. Thus, a detailed analysis is needed to balance accuracy, interpretability, and stability in system identification. 

The identified equations in Sec. \ref{equations_Ur5} demonstrate that, even in periodic regimes where both methods reproduce oscillatory behaviour well, WSINDy yields compact equations free of spurious small-amplitude terms, whereas SINDy equations include additional nonlinearities that may obscure physical interpretation and destabilise long-term simulations. However, when using WSINDy in a periodic regime, the library must be pruned to ensure stability, and the threshold is higher, enabling the identification of more interpretable dynamics. The RMSE for $\dot{y}$ is obtained using the predicted derivatives from the fitted model using \texttt{model.predict()} of the \texttt{pysindy} library, while the RMSE for $y$ is obtained by simulating the identified model over different lengths of time window. The shaded regions indicate the stability thresholds for the models. The figure highlights the trade-off between model sparsity, predictive accuracy, and simulation stability.

Figure~\ref{fig:integration_error} illustrates the relative error $\varepsilon(\%)$ in coefficient recovery as a function of the number of integration points $K$ used in the weak formulation. As shown in the plot, the error decreases significantly with increasing $K$, particularly for $K>500$, confirming the advantage of using a weak form to suppress noise. Beyond a certain threshold, the accuracy plateaus, indicating that an optimal number of integration points balances computational efficiency with noise robustness. In this study, we have used $K=1500$.

\begin{figure}[htbp]
    \centering
    \includegraphics[width=0.9\textwidth]{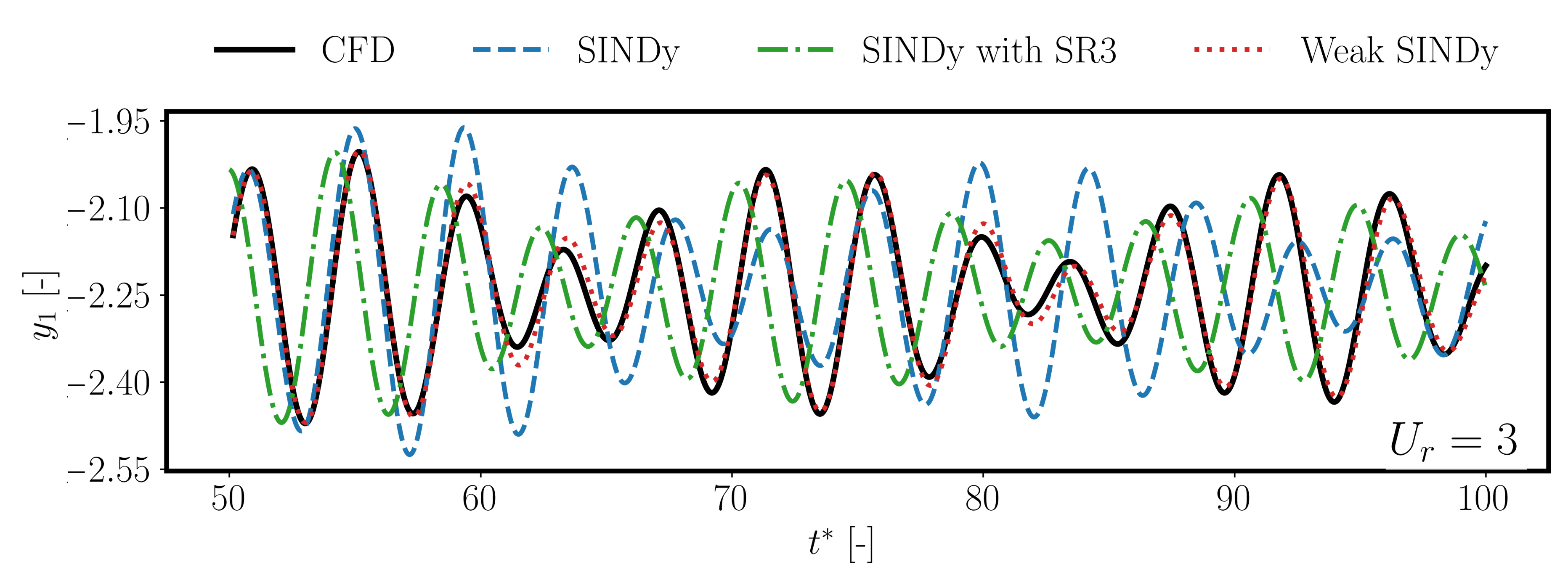}
    \caption{
    Comparison of the transverse displacement $y_1(t)$ at $U_r = 3$ obtained from CFD, SINDy identified dynamics with optimisers STLSQ and SR3, and the weak formulation of SINDy with optimiser SR3.
    }
    \label{fig:SR3}
\end{figure}

To further isolate the respective contributions of the optimisation strategy and the differentiation framework to identification accuracy in the aperiodic pre-lock-in regime, we present a comparison of the transverse displacement response $y_1(t)$ for three sparse identification configurations at $U_r = 3.0$: standard SINDy with the STLSQ optimiser, standard SINDy with the SR3 optimiser retaining strong-form differentiation, and WSINDy with SR3, in which both the differentiation framework and the optimiser are replaced simultaneously; see Fig.~\ref{fig:SR3}. In the present study, SINDy is consistently implemented using the STLSQ optimiser. However, in this figure, we also apply the SR3 optimiser to SINDy to clearly isolate the optimiser's effect. This allows us to distinguish between improvements arising purely from a change in optimisation strategy and those resulting from combining a different optimiser with a different differentiation framework. The standard SINDy-STLSQ model captures the overall oscillatory trend but exhibits noticeable amplitude and phase deviations from the CFD solution, primarily due to the sensitivity of finite-difference differentiation to oscillatory dynamics. When only the optimiser is changed to SR3 (while retaining the strong-form differentiation), the prediction improves, highlighting the contribution of SR3’s relaxed regularisation in producing more stable and consistent sparse coefficients. The best agreement with CFD is obtained with Weak-SINDy using SR3. By replacing pointwise differentiation with a weak (integral) formulation, numerical differentiation errors are significantly reduced. As a result, the model remains better aligned in phase and more accurate in amplitude over time. Overall, this comparison clearly separates the role of the optimiser from that of the differentiation strategy and demonstrates the combined benefit of SR3 optimisation and weak-form differentiation in more accurately capturing the underlying dynamics.

\section{SINDy-based sparse identification of POD temporal coefficients for near-wake flow reconstruction}
\label{sec:5}


\noindent Having demonstrated the capability of standard SINDy and WSINDy to recover physically consistent governing equations directly from the force and displacement time series of the coupled VIV system in Sec.~\ref{sec:4}, the present section extends the sparse identification framework to the full flow-field data, investigating whether SINDy can autonomously discover interpretable reduced-order representations of the spatiotemporal wake dynamics from the modal coefficient time series obtained via Proper Orthogonal Decomposition (POD) of the CFD velocity field.

\begin{figure}[htbp]
    \centering
    \includegraphics[width=0.75\textwidth]{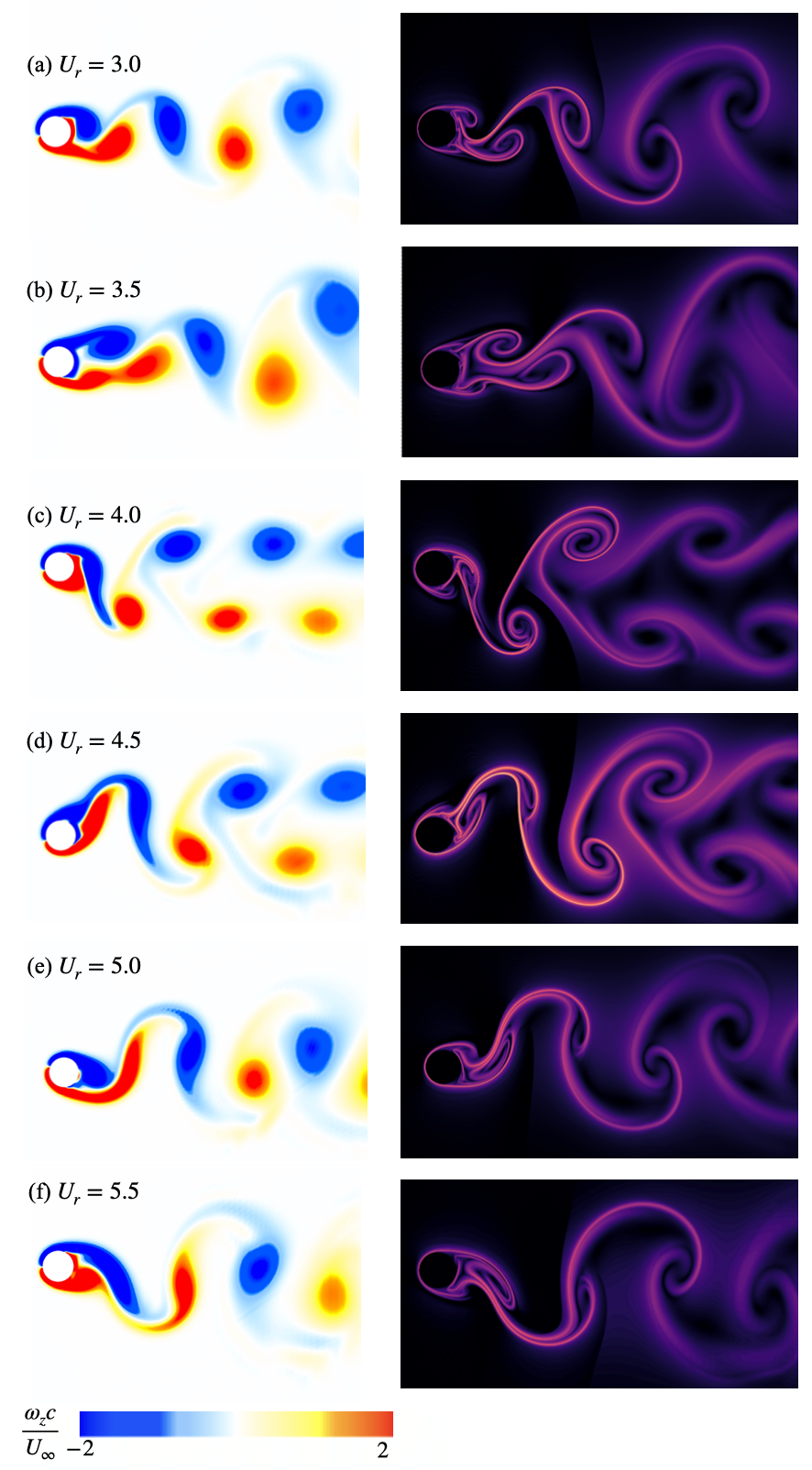}
    \caption{Instantaneous normalised spanwise vorticity contours $\omega_z c / U_\infty$ (left column) and corresponding Finite-Time Lyapunov Exponent (FTLE) fields (right column) for (a) $U_r = 3.0$, (b) $U_r = 3.5$, (c) $U_r = 4.0$, (d) $U_r = 4.5$, (e) $U_r = 5.0$, and (f) $U_r = 5.5$ at $t^* = 50$. Vorticity contours are presented for $\omega_z c / U_\infty \in [-2, 2]$; red and blue regions denote positive (anticlockwise) and negative (clockwise) vorticity, respectively.}
    \label{figure:FTLE}
\end{figure}

The instantaneous flow-field data obtained from the CFD simulations are examined through vorticity contours and Finite-Time Lyapunov Exponent (FTLE) fields at $t^* = 50$ to characterise the wake topology and identify the primary Lagrangian Coherent Structures (LCS) across the full range of reduced velocities considered; see Fig.~\ref{figure:FTLE}. FTLE ridges delineate the material surfaces that govern the organisation and transport of vortical structures in the cylinder wake, providing a measure of the local stretching rate of fluid particle trajectories. At $U_r = 3.0$, the cylinder displacement remains small, and the wake exhibits a compact recirculation region with relatively disorganised vortical structures, confirming that the system resides in the pre-lock-in regime. As $U_r$ increases to $3.5$, the LCS ridges reveal the emergence of larger, more coherent vortical structures and the beginnings of organised roll-up, consistent with the onset of quasi-periodic oscillations and the broad multi-frequency spectral content identified in the FFT spectra of Fig.~\ref{fig:CFD_timeseries_FFT}. The vorticity contours at this reduced velocity exhibit alternating positive and negative cores of increasing spatial extent, though the inter-vortex spacing remains irregular, indicative of the transitional character of the wake at the boundary of the pre-lock-in and lock-in regimes.

At $U_r = 4.0$, the wake transitions to a well-organised, periodic 2S vortex shedding pattern, characterised by regularly spaced, counter-rotating vortex cores and sharp, continuous FTLE ridges that extend far into the downstream wake. This case exhibits the most pronounced and geometrically regular LCS topology of all reduced velocities considered, consistent with strong fluid--structure synchronisation at lock-in onset and with the maximum oscillation amplitude reported by \citet{wang2017three} at this reduced velocity. The clarity of the FTLE ridges at $U_r = 4.0$ reflects a high degree of phase coherence between the vortex formation cycle and the cylinder motion, a feature that is directly relevant to the identifiability of the coupled system by the SINDy framework. As $U_r$ increases further to $4.5$ and $5.0$, the system remains within the lock-in regime, but the vortical structures become progressively larger and more spatially extended, reflecting the growing influence of structural displacement amplitude on wake organisation. At $U_r = 5.0$, the vorticity contours show well-separated, rapidly convecting vortex cores with clear alternation, and the FTLE ridges remain continuous and well-defined, indicative of sustained phase coherence between structural oscillation and vortex formation throughout the lock-in regime. At $U_r = 5.5$, the LCS topology reveals larger, more elongated vortical structures with increased inter-vortex spacing and a reduced shedding frequency, consistent with the downward shift in dominant frequency observed in the FFT spectra and indicative of the system approaching the upper boundary of the lock-in regime. Collectively, the vorticity and FTLE visualisations confirm that the CFD simulations reproduce the canonical pre-lock-in, lock-in, and post-lock-in wake-topology transitions associated with VIV, and that the flow-field data exhibit the physical characteristics, ranging from disorganised pre-lock-in shedding to strongly periodic lock-in dynamics, that are necessary to provide a rigorous and varied test bed for the subsequent sparse identification and POD analyses presented in Sections~\ref{sec:4} and~\ref{sec:5}.

The detailed description of the snapshot-based POD analysis adopted in this study is presented in \ref{app:C}. Having established the POD basis and confirmed that ten modes provide a sufficiently reduced-order representation of the dominant wake dynamics across the three reduced velocities considered, the scaled temporal coefficients $\tilde{a} = \Sigma V^\mathrm{T}$ are used as the state variables for sparse identification. Both SINDy and WSINDy are trained on temporal coefficient trajectories extracted from CFD snapshot data over $t^* \in [50, 150]$ and subsequently integrated forward in time to assess their ability to reconstruct the dominant spatiotemporal flow structures. The identification is performed using a library of first-degree polynomial candidates; the rationale for this choice and the challenges associated with higher-degree alternatives are discussed in detail below. It is emphasised that the identification problem addressed in this section is fundamentally more challenging than the two-variable structural--wake formulation of Sec.~\ref{sec:4}: the ten-dimensional state space, the irregular temporal behaviour of the higher-order modal coefficients, and the absence of strong physical constraints on inter-modal coupling collectively represent a significantly more demanding test of the sparse identification framework. The spatial modal structures and the corresponding temporal coefficient reconstructions for $U_r = 3.0$, $3.5$, and $4.0$ are presented in the following figures, alongside a direct comparison of the flow fields reconstructed from the SINDy- and WSINDy-identified models against the reference CFD snapshots.

\begin{figure}[htbp]
    \centering
     \includegraphics[width=0.8\textwidth]{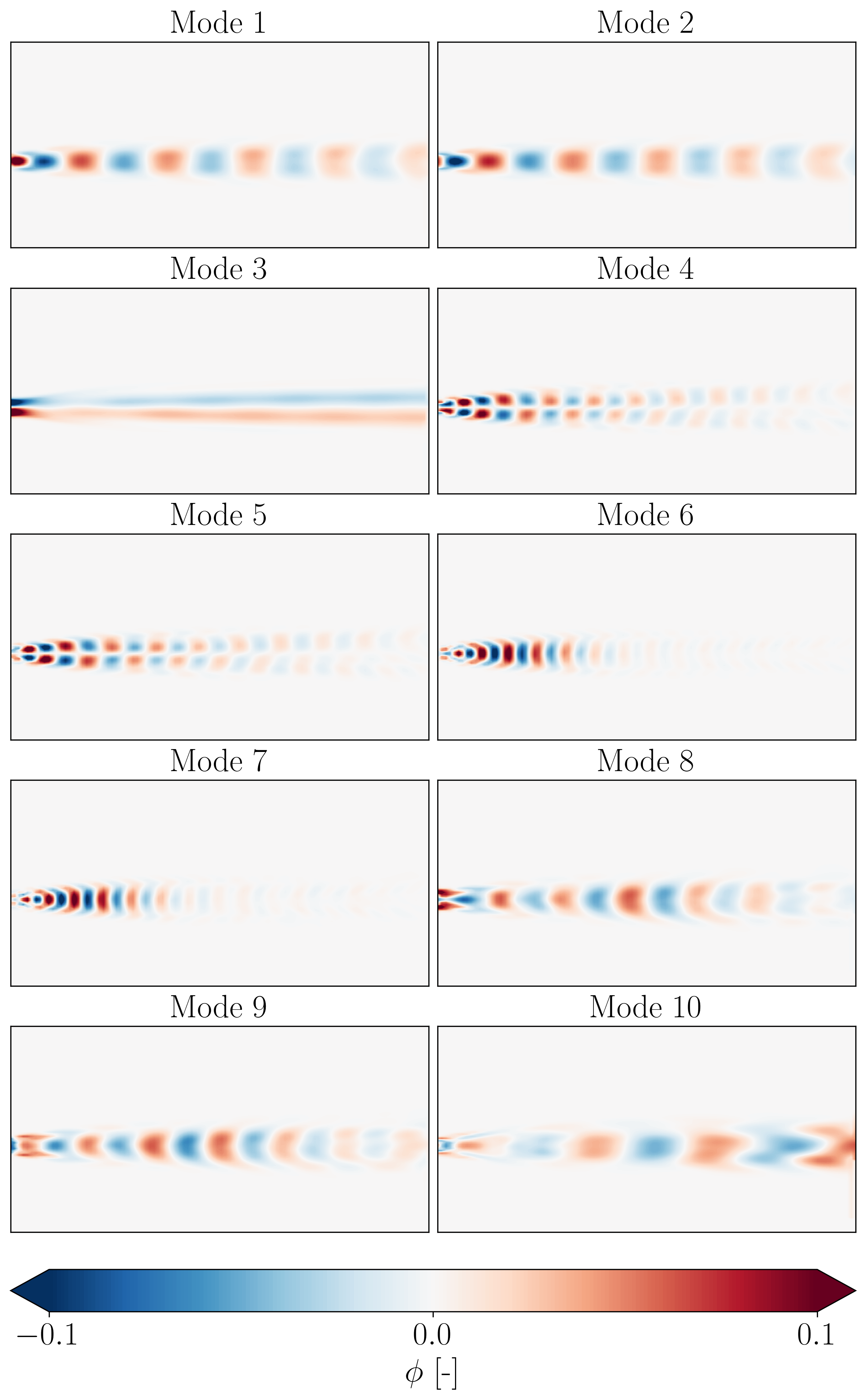}
    \caption{First 10 high-energy spatial POD modes of the unsteady wake downstream the cylinder for $ U_r = 3 $.}
    \label{fig:modes_ur3}
\end{figure}

\begin{figure}[htbp]
    \centering
    \includegraphics[width=0.8\textwidth]{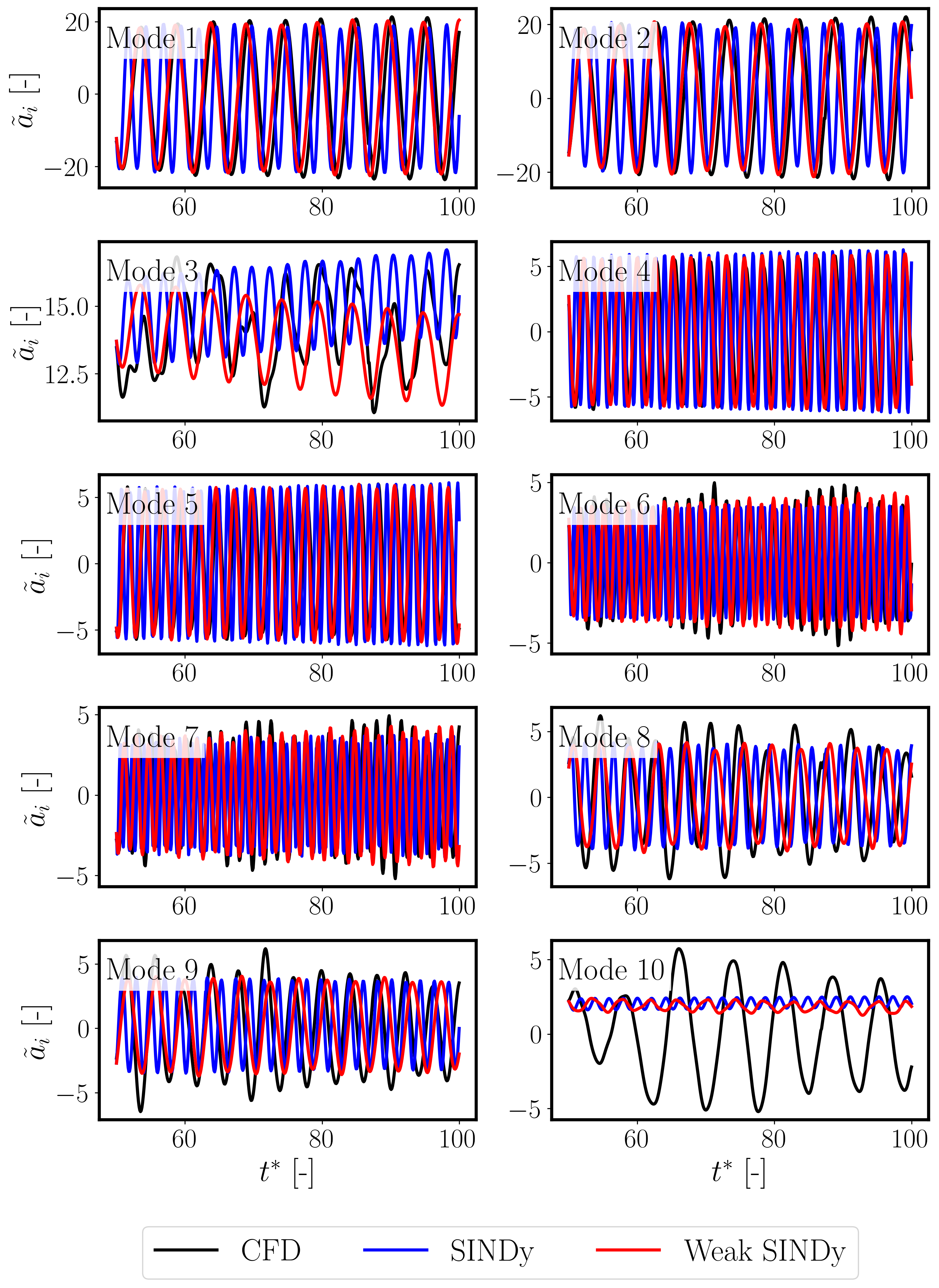}
    \caption{Comparison of $\tilde{a}$ predicted by SINDy, WSINDy, and the original trajectory for $ U_r = 3 $. Here, $\tilde{a} = \Sigma V^{\mathrm{T}}$, where $\Sigma$ denotes the diagonal matrix of singular values and $V^{\mathrm{T}}$ represents the temporal coefficients.}

    \label{fig:temporalcoeffmodes_ur3}
\end{figure}

\begin{figure}[htbp]
    \centering
    \includegraphics[width=0.8\textwidth]{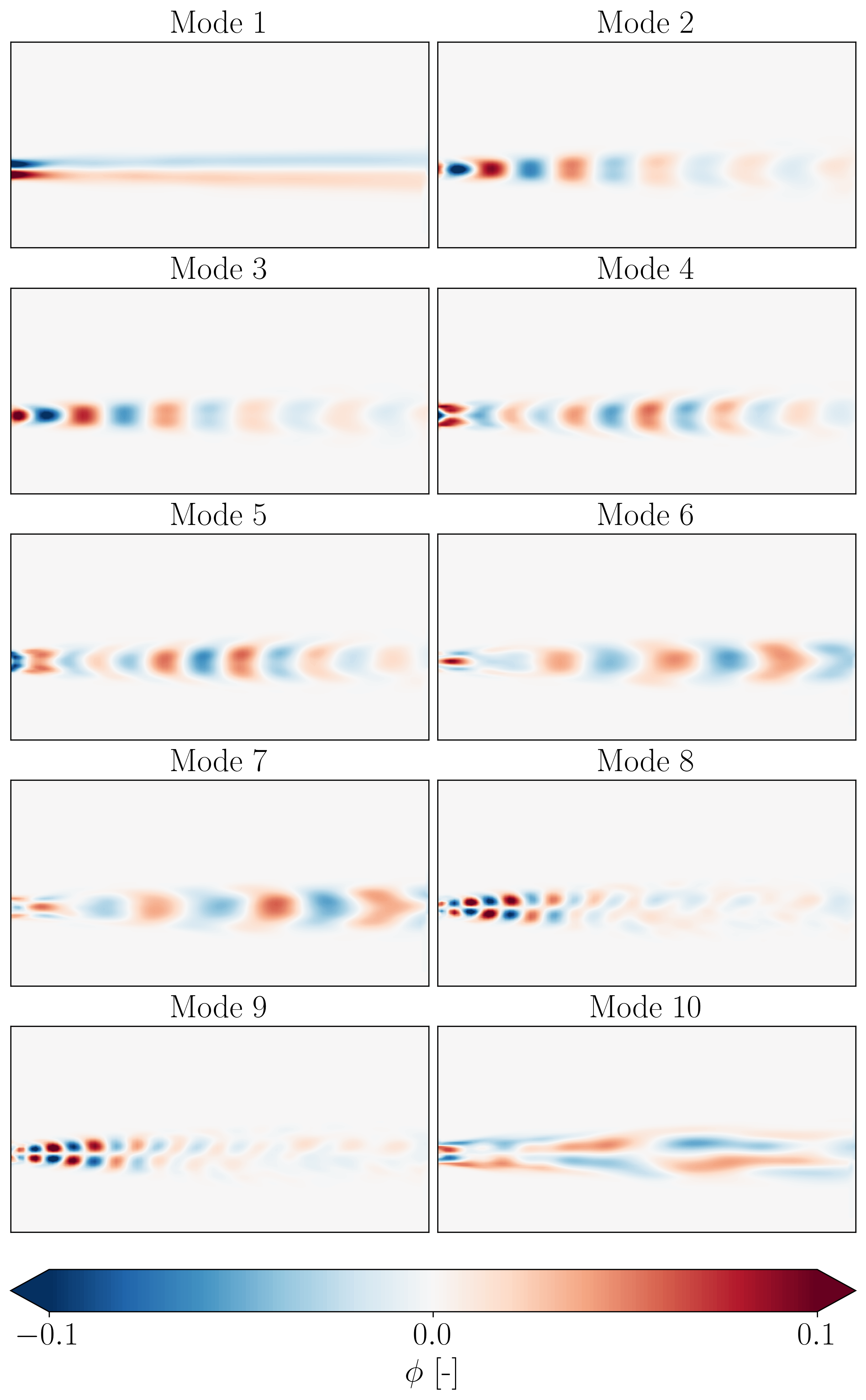}
    \caption{First 10 high-energy spatial POD modes of the unsteady wake downstream the cylinder for $ U_r = 3.5$.}
    \label{fig:modes_ur3.5}
\end{figure}

\begin{figure}[htbp]
    \centering
    \includegraphics[width=0.8\textwidth]{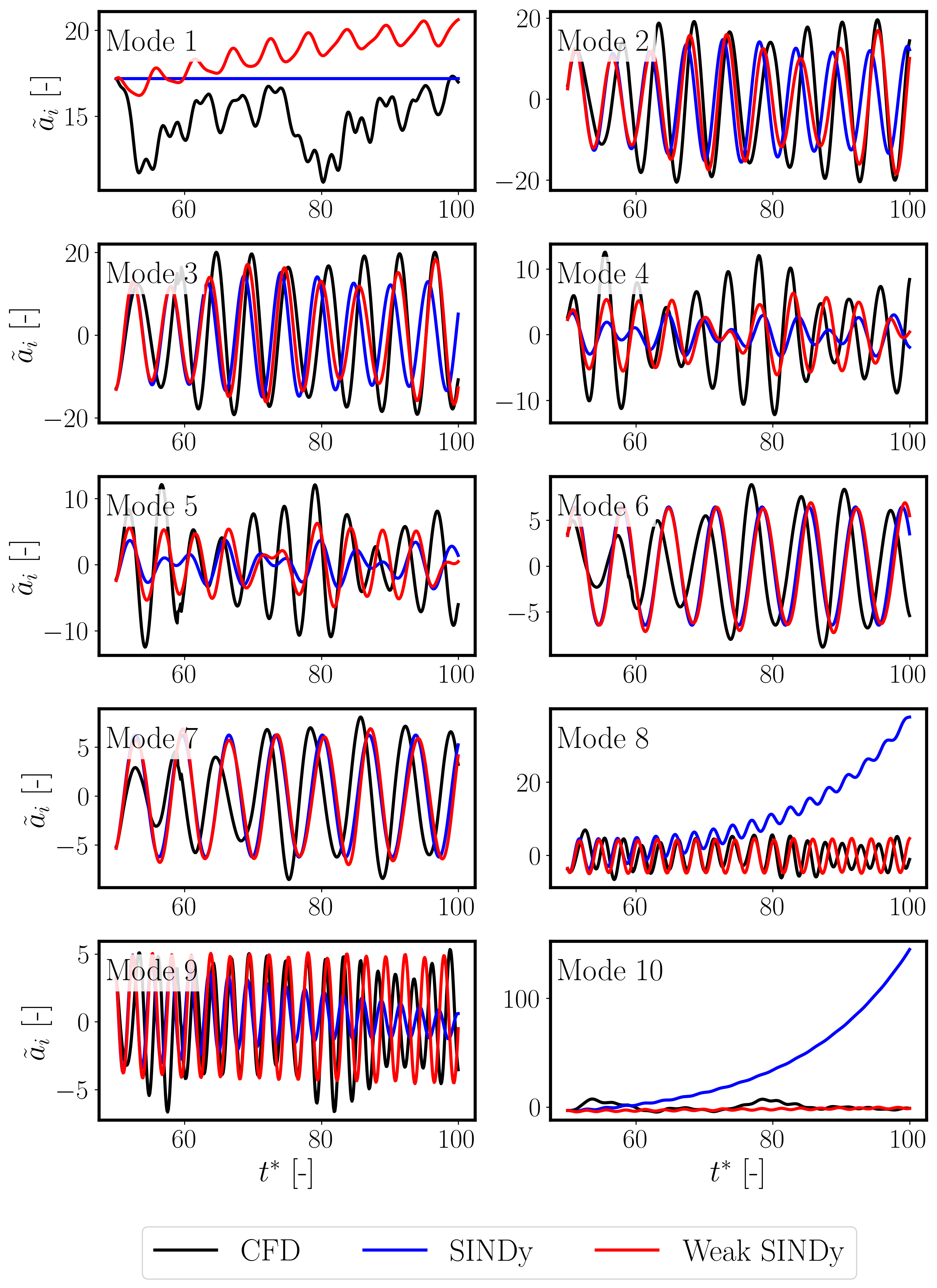}
    \caption{Comparison of $\tilde{a}$ predicted by SINDy, WSINDy, and the original trajectory for $ U_r = 3.5 $.}
    \label{fig:temporalcoeffmodes_ur3.5}
\end{figure}

\begin{figure}[htbp]
    \centering
    \includegraphics[width=0.8\textwidth]{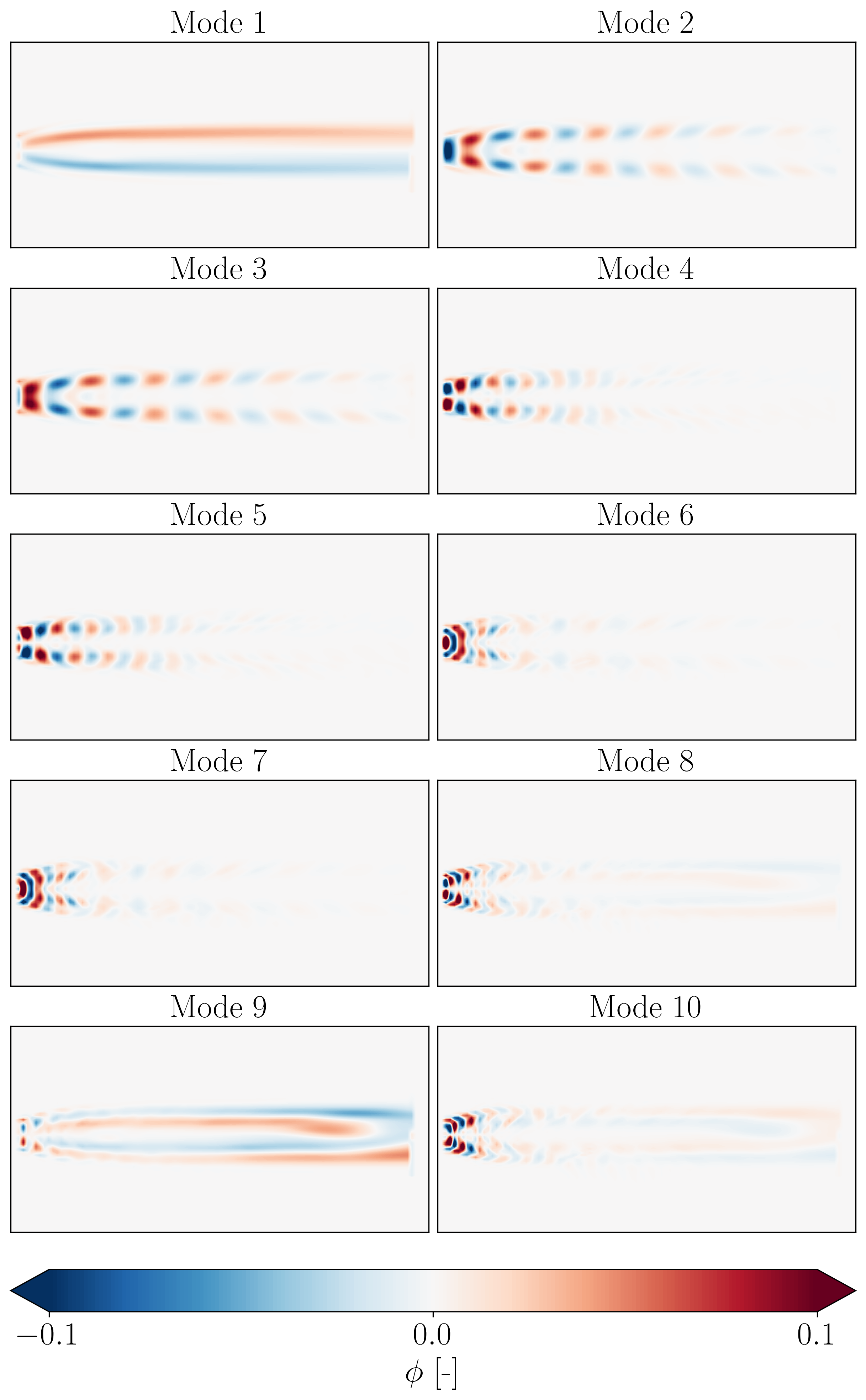}
    \caption{First 10 high-energy spatial POD modes of the unsteady wake downstream the cylinder for $ U_r = 4$.}
    \label{fig:modes_ur4}
\end{figure}

\begin{figure}[htbp]
    \centering
    \includegraphics[width=0.8\textwidth]{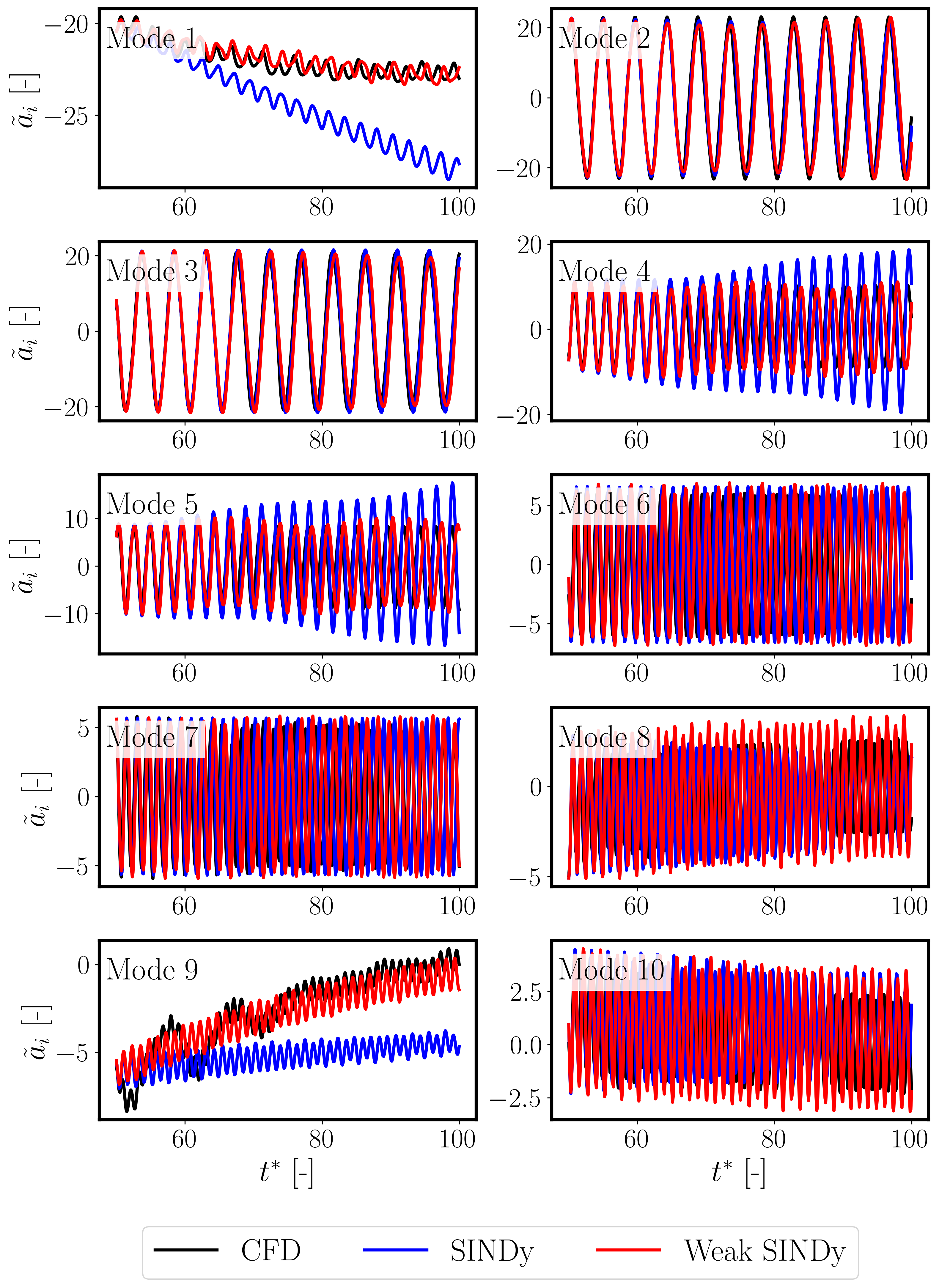}
    \caption{Comparison of $\tilde{a}$ predicted by SINDy, WSINDy, and the original trajectory for $U_r = 4$.}

    \label{fig:temporalcoeffmodes_ur4}
\end{figure}

Figures~\ref{fig:modes_ur3}, \ref{fig:modes_ur3.5}, and \ref{fig:modes_ur4} present the first ten spatial POD modes for $U_r = 3.0$, $3.5$, and $4.0$, respectively, for direct visual comparison of modal shapes across reduced velocities. Figures~\ref{fig:temporalcoeffmodes_ur3}, \ref{fig:temporalcoeffmodes_ur3.5}, and \ref{fig:temporalcoeffmodes_ur4} display the original temporal coefficients $\tilde{a} = \Sigma V^\mathrm{T}$ for modes 1--10 alongside those recovered by SINDy and WSINDy. Whilst both methods exhibit substantially lower reconstruction accuracy for the ten-mode system than for the two-variable structural--wake formulation of Section~\ref{sec:4}, WSINDy consistently produces temporal coefficient trajectories that more closely follow the original POD coefficients across all three reduced velocities, particularly for the lower-order modes that carry the majority of the flow energy. This relative advantage is most pronounced at $U_r = 3.0$ and $3.5$, where the aperiodic, broadband character of the pre-lock-in dynamics renders the pointwise derivative estimates used by standard SINDy especially susceptible to noise-induced bias, whereas WSINDy's integral projection formulation partially mitigates this sensitivity. Nonetheless, neither method achieves quantitatively reliable reconstruction of the higher-order modal coefficients, and both identified systems exhibit phase drift and amplitude errors that grow with modal index.

The primary source of this limitation is the use of a first-degree polynomial candidate library for the ten-variable system, which yields only ten candidate terms and, by construction, cannot represent the quadratic and higher-order inter-modal interactions present in the true nonlinear wake dynamics. Extending the library to second- or third-degree polynomials would, in principle, provide the representational capacity to capture such interactions---a degree-2 library for ten variables yields 66 candidate terms and a degree-3 library yields 286---but in practice this dramatically exacerbates the model instability problem already encountered in Section~\ref{sec:4}. The combination of a large candidate library, noisy higher-order modal coefficients, and the absence of strong physical constraints on inter-modal coupling consistently produces unstable identified systems across all sparsity thresholds tested, and the use of a linear library therefore represents a deliberate compromise between representational completeness and numerical stability.


\begin{figure}[htbp]
    \centering
    \includegraphics[width=1\textwidth]{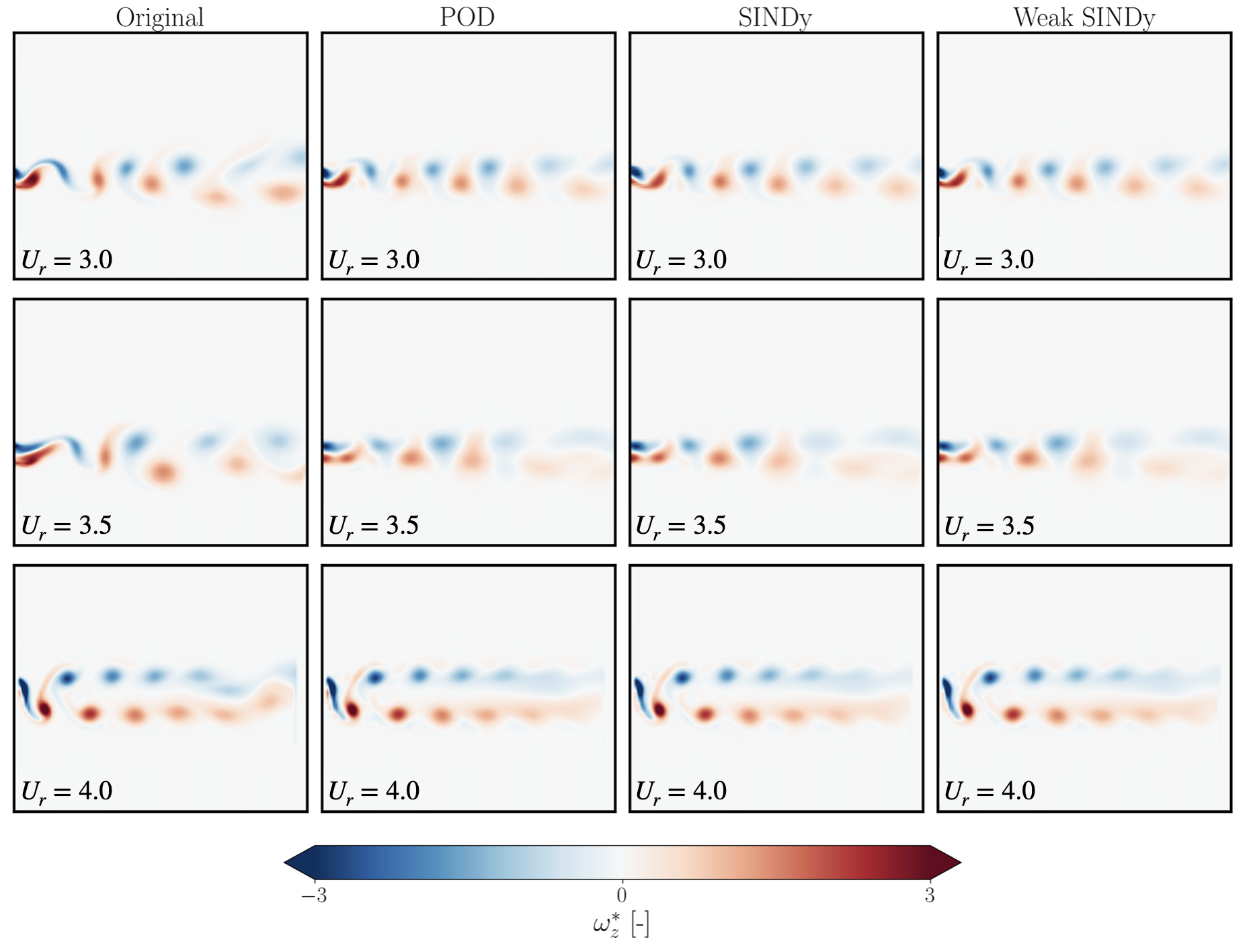}
    \caption{Comparison of wake patterns for different reduced-order modelling approaches across two reduced velocity values, $U_r = 3$, $U_r = 3.5$ and $U_r = 4$. }
    \label{fig:pod_sindy_comparison}
\end{figure}

The reconstructed flow fields for all three reduced velocities are presented in Figure~\ref{fig:pod_sindy_comparison}, with each snapshot selected from an early instance of the respective regime. Both methods qualitatively reproduce the dominant large-scale flow structures, and WSINDy produces visually closer agreement with the reference CFD fields, consistent with its superior temporal coefficient reconstruction. However, the quantitative agreement between both methods and the original data remains limited, and discrepancies are most pronounced in the near-wake region immediately behind the cylinder, where spatial gradients are steepest, and the contribution of higher-order, less accurately identified modes is greatest. Errors accumulate progressively during forward-time simulation as small discrepancies in the identified modal dynamics compound across time steps, and this accumulation is more severe for standard SINDy than for WSINDy. These results confirm that whilst POD--SINDy is capable of capturing the qualitative topology of the wake, and whilst WSINDy offers a measurable improvement over standard SINDy under the present noisy, high-dimensional conditions, achieving quantitatively reliable flow-field reconstruction from CFD data with a linear candidate library remains beyond the reach of the present identification framework.

\section{Physical interpretability and structural consistency of identified sparse models}
\label{sec:6}

All identified SINDy models, whether derived from the mathematical wake-oscillator model or from CFD data, successfully capture the key dynamical signatures of VIV, including lock-in, amplitude modulation, and the phase relationship between structural displacement and wake forcing, with varying accuracy across $U_r$ regimes. However, trajectory-tracking accuracy alone is an insufficient criterion for evaluating the physical validity of a sparse identified model; the sparsity, term consistency, and physical interpretability of the identified equations are equally important indicators of model quality and generalisability. The following discussion therefore compares the four identified models of flow past an elastically mounted cylinder, namely, the original wake-oscillator mathematical model, SINDy applied to mathematical-model trajectories, SINDy applied to CFD data, and WSINDy applied to CFD data, according to their interpretability, noise robustness, and structural fidelity. Representative equation sets at $U_r = 5$ are presented in \ref{equations_Ur5} to provide a concrete illustration of the differences between the four models.

When applied to clean trajectory data generated by the wake-oscillator model of~\ref{app:A}, SINDy recovers the governing dynamics with high-fidelity using a full third-degree polynomial library with a sparsity threshold of $\lambda = 0.1$. The identified equations preserve the structural organisation of the original model, including the linear inertial and damping terms in the structural equation and the nonlinear Van der Pol-type term $(q_1^2 - 1)\dot{q}_1$ in the wake equation, and yield excellent numerical accuracy for the recovered coefficients. This result confirms that, under idealised conditions of clean, densely sampled data, SINDy can recover not only the qualitative structure but also the precise parametric form of the underlying physical model.

When applied to CFD-derived data, the identification landscape changes substantially. Standard SINDy produces significantly denser equation sets, particularly in the wake dynamics ($\dot{q}_1$, $\dot{q}_2$), where a wide variety of high-order nonlinear and mixed interaction terms, including $y_1 q_1$, $q_1^2$, and $y_2^2$, appear with non-negligible coefficients despite having limited physical justification within the wake-oscillator paradigm. This term proliferation can be attributed primarily to the noise present in the CFD-derived lift force signal: unlike the structural displacement $y_1$, which varies smoothly, the lift coefficient exhibits high-frequency fluctuations that standard SINDy, operating on pointwise derivative estimates, misinterprets as genuine dynamical content and compensates for by recruiting additional library terms. Although the resulting model achieves accurate trajectory tracking, it does so by including physically unwarranted compensatory terms rather than by recovering the true governing structure, a distinction of considerable practical importance for extrapolative prediction and physical interpretation.

WSINDy, applied to the same CFD data, produces markedly sparser and more physically coherent equation sets, particularly in the wake equations. The retained terms are few in number and consistent with the expected wake-oscillator structure: linear coupling terms and specific quadratic interactions are preserved, whilst the spurious high-order terms that proliferate in the standard SINDy models are suppressed. This behaviour is a direct consequence of WSINDy's integral projection formulation, which distributes the effect of high-frequency noise across compactly supported test functions rather than allowing it to corrupt pointwise derivative estimates, and confirms the suitability of the weak-form approach for sparse identification from real-world or CFD-derived data. Note, however, that model stability in WSINDy required the manual exclusion of the $q_2^2$ term from the candidate library prior to identification, whereas standard SINDy could discover stable dynamics from the full second-degree polynomial library. This distinction highlights a practical limitation of WSINDy in the present context: whilst its integral formulation confers superior noise robustness, it may in some cases require user intervention to enforce stability, introducing a degree of a priori structural knowledge that partially undermines the fully data-driven character of the identification.

Two further structural observations merit discussion. First, the kinematic identity $\dot{y}_1 = y_2$ is recovered consistently and correctly across all four models, confirming that this relationship is sufficiently dominant in the data to be robustly identified even under noisy conditions. By contrast, the analogous wake identity $\dot{q}_1 = q_2$, which is present in the mathematical model by construction, is recovered only by SINDy applied to the mathematical-model trajectories and is absent from both CFD-based models. This asymmetry reflects the greater susceptibility of the wake variable $q_1$ to noise-induced effects than that of the structural displacement $y_1$, underscoring the difficulty of reliably identifying the kinematic structure of the wake dynamics from noisy data. Second, nonlinear cubic terms, which carry genuine physical significance within the Van der Pol wake-oscillator framework, are present only in the model identified from mathematical-model data and are absent from all CFD-based models. Their absence does not necessarily imply that such dynamics are absent from the true system; rather, it reflects the practical difficulty of reliably identifying high-order terms from noisy CFD data, where the inclusion of cubic library terms consistently destabilises the identified system and the signal-to-noise ratio is insufficient to distinguish genuine cubic dynamics from noise artefacts.

\section{Concluding remarks}
\label{sec:7}

This study presents the first systematic comparative assessment of SINDy and its weak-formulation variant (WSINDy) for data-driven discovery of governing equations in vortex-induced vibration, conducted on synthetic wake-oscillator data and high-fidelity CFD data spanning the pre-lock-in, lock-in, and post-lock-in regimes.

When applied to clean wake-oscillator trajectories, SINDy recovers the governing equations with high fidelity, preserving both the qualitative structure and numerical coefficients of the original model, including the nonlinear Van der Pol-type term and linear structural contributions, thereby confirming that sparse regression can serve as a fully autonomous equation-discovery tool under idealised conditions. On the other hand, when applied to CFD data, standard SINDy accurately tracks the periodic lock-in response but produces dense, physically non-interpretable equation sets under aperiodic pre-lock-in conditions, where noise in the lift force signal affects pointwise derivative estimates. WSINDy consistently produces sparser, more physically coherent models across all reduced velocities, with its integral-projection formulation acting as a natural low-pass filter that suppresses noise-induced proliferation of terms. This advantage is most pronounced in the pre-lock-in regime, confirming the suitability of the weak-form approach for identification even from an aperiodic response.

Extension to full near-wake flow-field reconstruction via POD--SINDy revealed a fundamental balance among modal dimensionality, library expressiveness, and numerical stability. A linear candidate library with ten POD modes produced qualitatively reasonable but quantitatively limited reconstructions; WSINDy again outperformed standard SINDy. Extending the library to degree-2 or degree-3 polynomials (66 and 286 terms, respectively) consistently yielded unstable systems, confirming that the principal barrier is not library expressiveness per se, but noise sensitivity and high-dimensional system instability.

Overall, the present findings demonstrate that WSINDy offers a measurable advantage over standard SINDy for discovering VIV equations from CFD data, providing a more favourable balance among trajectory accuracy, physical interpretability, and noise robustness. Future work should prioritise noise-aware sparse regression formulations, alongside physics-informed regularisation to constrain identification to physically realisable structures without manual pruning. Extending the POD basis beyond ten modes with symmetry-informed sparse libraries offers a natural route to improved flow-field reconstruction. In the longer term, application to tandem-cylinder arrangements and three-dimensional geometries at higher Reynolds numbers will provide a more demanding test of the robustness and generalisability of weak-form sparse identification for engineering fluid--structure interaction problems.

\appendix

\section{Coupled wake-oscillator model: methodology and validation}
\label{app:A}

\begin{figure}[htbp]
    \centering
    \includegraphics[width=0.55\textwidth]{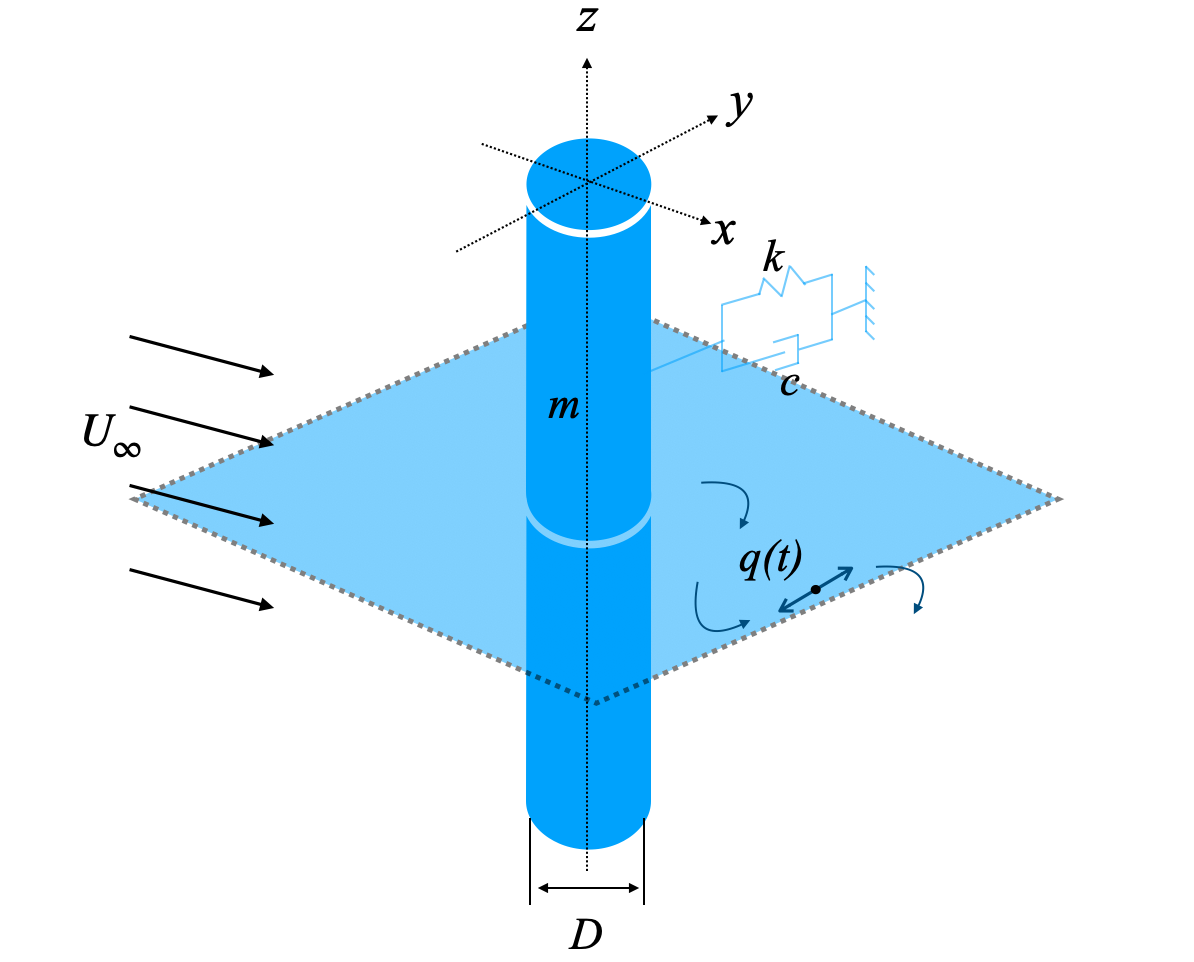}
    \caption{Schematic representation of the VIV system, illustrating the coupled interaction between the 1-DoF structural oscillator and the wake oscillator, representative of the vortex-induced oscillatory force along the cross-flow direction.}
    \label{fig:VIVsinglemodel}
\end{figure}

In this approach, the VIV system is modelled using a phenomenological model through a coupled system of two second-order ordinary differential equations (ODEs). The first describes a linear oscillator, modelling the motion of a circular cylinder with a single-degree-of-freedom (1-DoF) in the cross-flow direction, whereas the second corresponds to the non-linear Van der Pol oscillator equation, used to capture wake dynamics. The 1-DoF VIV system is schematically represented in Fig.~\ref{fig:VIVsinglemodel}.

\noindent The cross-flow motion of the cylinder is modelled as a linear oscillator, given by, 

\begin{equation}
m \ddot{Y} + c \dot{Y} + k Y = S,
\end{equation}
\noindent where $Y$ denotes the structural displacement, $m =  m_s + m_f$ is the total mass, where $m_s$ and $m_f$ are the structural and fluid masses, respectively. $c$ is the damping coefficient, $k$ is the spring constant, and $S$ is the forcing term due to vortex shedding. The ODE is non-dimensionalised by introducing the dimensionless time $t = \Omega_f T$ and dimensionless displacement $y = \frac{Y}{D}$, which yields:

\begin{equation}
\ddot{y} + \left(2\xi \delta + \frac{\gamma}{\mu}\right) \dot{y} + \delta^2 y = s,
\label{3}
\end{equation}
\noindent where \( \delta = \frac{\Omega_s}{\Omega_f} \) is the reduced angular frequency (the ratio of structural angular frequency \(\Omega_s \) to vortex shedding angular frequency \(\Omega_f \)), \(\gamma \) is the stall parameter,  $\xi$ is the damping ratio, \( s = \frac{S}{mD \Omega_f^2} \) is the dimensionless coupling term, $D$ is the diameter of the cylinder, and $T$ is the dimensional time.

The Van der Pol equation is used to model the periodic behaviour of the lift through the wake parameter $q$, and is defined as follows:
\begin{equation}
\ddot{q} + \epsilon \Omega_f (q^2 - 1) \dot{q} + \Omega_f^2 q = \frac{F}{D},
\end{equation}
\noindent where $q$ is the wake oscillator variable; $\epsilon$ is the parameter representing nonlinearity, and $F$ is the forcing term associated with the structural response. In the non-dimensional form, considering \( t = \Omega_f T \), the governing equation becomes: 
\begin{equation}
\ddot{q} + \epsilon (q^2 - 1)\dot{q} + q = f,
\label{4}
\end{equation}
\noindent where \( f = \tfrac{F}{D \Omega_f^2} \) denotes the dimensionless coupling term.


The coupling terms, $s = s(q)$ and $f = f(y)$, describe the interaction between structural vibrations and wake dynamics. The structural coupling term $s$, representing the effect of vortex forces on the cylinder, is expressed as: $s = \frac{S}{mD \Omega_f^2} = \frac{SD}{4m \pi^2 St^2 U_\infty^2}$. The wake coupling term $f$, representing the effect of structural motion on the wake, is given by $f = \frac{F}{D \Omega_f^2} = \frac{F D}{4 \pi^2 St^2 U_\infty^2}$. The reduced angular frequency is defined as $\delta = \frac{\Omega_s}{\Omega_f} = \frac{D \Omega_s}{2 \pi St U_\infty} = \frac{1}{St} \left(\frac{D \Omega_s}{2 \pi U_\infty}\right) = \frac{1}{St U_r}$, where $U_r = \frac{2 \pi U_\infty}{\Omega_s D}$ is the reduced flow velocity, $St$ is the Strouhal number, and $U_\infty$ is the free-stream velocity.

These non-dimensional governing equations (Eqs.~\ref{3} and \ref{4}) are transformed into a system of four first-order ODEs (considering $y = y_1, \dot{y} = y_2, q = q_1, \text{and} \ \dot{q} = q_2$) that represent the state-space of the dynamical system; see Eqs. \eqref{16}-\eqref{19}. The system of dimensionless equations is numerically solved using the Runge-Kutta method over a range of reduced velocities from $1$ to $20$ to generate training data for the SINDy algorithm. The system is initialised with \( y_1(0) = 0 \), \( y_2(0) = 0 \), \( q_1(0) = 1 \) and \( q_2(0) = 1 \). 


\begin{equation}
\label{16}
\dot{y}_1 = y_2,
\end{equation}

\begin{equation}
\label{17}
\dot{y}_2 = s - \left(2\xi \delta + \frac{\gamma}{\mu}\right) y_2 - \delta^2 y_1,
\end{equation}

\begin{equation}
\label{18}
\dot{q}_1 = q_2,
\end{equation}

\begin{equation}
\label{19}
\dot{q}_2 = f - \epsilon (q_1^2 - 1) q_2 - q_1.
\end{equation}

Two possible forms of the coupling term, based on acceleration and velocity \citep{MainPaper}, are considered to assess the accuracy of VIV representation when compared with reference data from \citet{govardhan2000modes}. In acceleration coupling, the coupling term is calculated as $f = A\ddot{y} = A\dot{y_2} = A(s  - (2\xi\delta + \frac{\gamma}{\mu})y_2 - \delta^2 y_1).$ On the other hand, in velocity coupling, it is taken as $f = A\dot{y} = Ay_2$. Here, $A$ is a coupling constant.

The parameters of the ODE system are determined from experimental data on free and forced oscillations of circular cylinders \citep{MainPaper}. The values are given as: $\xi = 0.0052$, $\delta = \frac{1}{St \, U_r} = \frac{1}{0.2 \, U_r}$, where $U_r$ is the reduced velocity, and $St = 0.2$ is assumed for the sub-critical Reynolds number range ($300 \leq Re \leq 1.5 \times 10^5$). The stall parameter is given by $\gamma = \frac{C_D}{4 \pi St} = \frac{2}{4 \pi (0.2)} \approx 0.8$, where \(C_D\) is the amplified drag coefficient. The mass ratio is defined as $m^* = \frac{4}{\pi} \mu - C_M = \frac{0.2}{\pi M} - C_M \approx 0.52$. Here, $C_M = 1$ is the constant added-mass coefficient, derived from potential flow theory for a circular cross-section and $M = \frac{0.05}{\mu} \approx 0.0419$ is the non-dimensional mass number. The parameters $\epsilon = 0.3$ and $A = 40\epsilon$ are chosen based on experimental observations for better prediction.

\begin{figure}[htbp]
    \centering
    \includegraphics[width=0.6\textwidth]{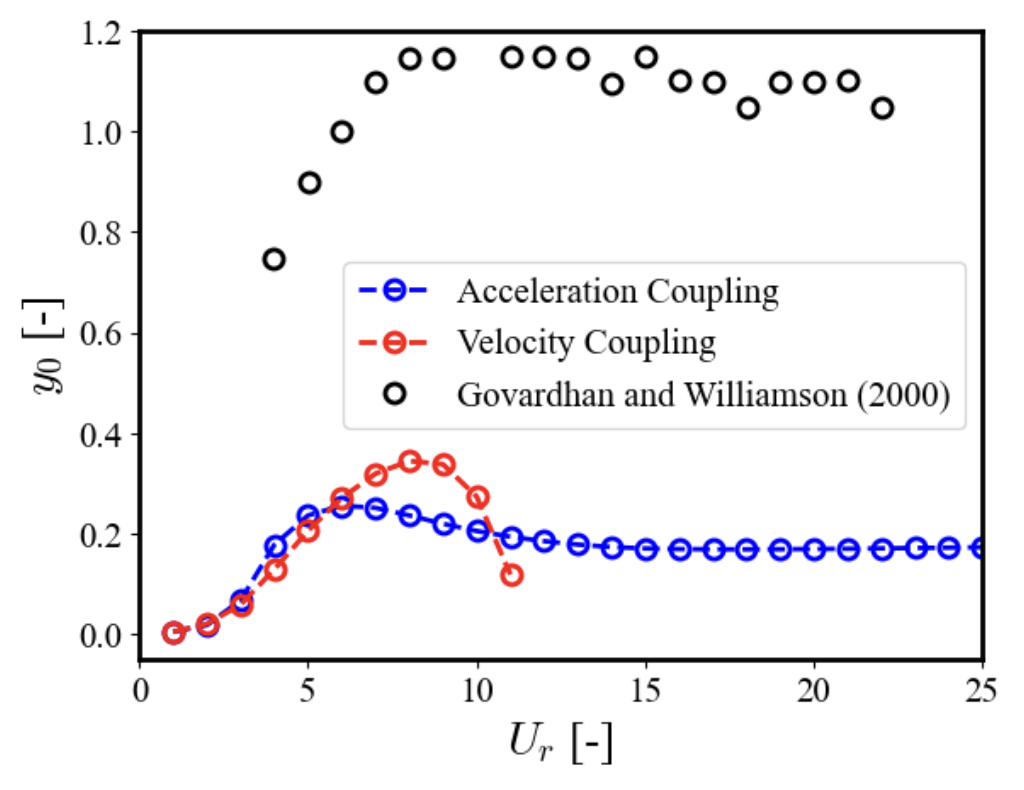}
    \caption{The dimensionless structural displacement amplitude $y_0$ obtained from the coupled wake-oscillator model for varying reduced velocity $U_r$, compared with reference experimental results obtained from \citet{govardhan2000modes}. The present results corroborates with those reported by \citet{Facchinetti}.}
    \label{figure: plotURy01}
\end{figure}

Figure \ref{figure: plotURy01} presents the variation of non-dimensional structural displacement amplitude $y_0$ with reduced velocity $U_r$. This low-order model reproduces the lock-in behaviour when the acceleration coupling term is included, in agreement with experimental observations by \citet{govardhan2000modes}. This confirms that acceleration coupling provides a more reliable representation of lock-in than the velocity coupling, a key feature of VIV. However, the model underestimates $y_0$, indicating the need for enhanced modelling strategies to more accurately capture the amplitude growth during lock-in.

\section{CFD simulations: methodology, verification, and validation}
\label{app:B}

\noindent An elastically mounted 1-DoF circular cylinder of diameter $D$ is 
considered, with simulations performed across a range of reduced 
velocities spanning the pre-lock-in, lock-in, and post-lock-in regimes, 
and the resulting force and displacement time series are subsequently 
provided as input to the sparse identification algorithm. The laminar incompressible fluid flow is governed by the Navier-Stokes equations:

\begin{equation}
     {\nabla }  \cdot {\bf{u}} = 0,
\end{equation}
\begin{equation}
\frac{\partial {\bf u}}{\partial t} + ( {\bf u} \cdot {\nabla}){ \bf u} = -\frac{1}{\rho_f}{\nabla}p + \nu{\nabla}^2 \bf{u}. \label{eq:NavierStockes} \end{equation}
Here, $\bf{u}$ denotes the fluid velocity vector, $\rho_f$ is the fluid density, $p$ is the fluid pressure, and $\nu$ is the kinematic viscosity of the fluid. 

The structural counterpart is modelled as a 1-DoF elastically supported circular cylinder, and the corresponding equation of motion can be given by

\begin{equation}
m_s \ddot{Y} + c \dot{Y} + k Y = F_y,
\end{equation}

\noindent where $Y$ denotes the structural displacement, $m$ is the structural mass, $c$ is the damping coefficient, $k$ is the spring constant, and $F_y$ is the transverse lift force. The Reynolds number (based on the uniform free-stream and the cylinder diameter as the velocity scale and the length scale, respectively) is set to $Re=U_\infty D / \nu = 150$ and the mass ratio, defined as $ m^* = \frac{4 m_s}{\rho_f D^2 H \pi}$, is taken as 2.546 following \citet{wang2017three}. Here, $m_s$ is the structural mass, $H$ is the height of the cylinder.
The reduced velocity is defined as follows: $U_r = \frac{U_\infty}{f_n D}$, where $f_n$ is the natural frequency of the structure.

In this study, simulations are conducted using the open-source finite-volume CFD code {\tt OpenFOAM}. The {\tt pimpleFoam} solver is used, which is based on the PIMPLE algorithm, a combination of the Pressure-Implicit with Splitting of Operators (PISO) and Semi-Implicit Method for Pressure-Linked Equations (SIMPLE) algorithms. Ten outer corrector loops and one pressure corrector loop are included, enabling efficient handling of transient incompressible flows. Temporal discretisation is performed using a second-order backward differencing scheme. Spatial discretisation is carried out using a second-order accurate Gauss linear scheme for gradients, a Gauss cubic interpolation scheme for divergence terms and a Gauss linear corrected scheme for Laplacian terms. Linear interpolation with a corrected surface-normal gradient scheme is adopted. The pressure equation is solved using a Geometric-Algebraic MultiGrid (GAMG) solver with a Gauss-Seidel smoother. Velocity is solved using a smooth solver with a symGaussSeidel smoother. Grid cell motion is handled via a preconditioned conjugate gradient (PCG) solver with a diagonal incomplete Cholesky (DIC) preconditioner. The rigid-body motion is solved using the {\tt sixDoFRigidBodyMotion} solver in {\tt OpenFOAM} with a {\tt symplectic} time-integration scheme with the values of the {\tt accelerationRelaxation} and {\tt accelerationDamping} parameters as 0.9 for faster convergence. Although the solver allows for six degrees of freedom, the motion is constrained to a single translational degree of freedom in the cross-flow direction. A partitioned weak-coupling approach is adopted for fluid-structure interaction, in which the flow field and the structural response are solved at each time step. 

\begin{figure}[htbp]
    \centering
     \includegraphics[width=1\textwidth]{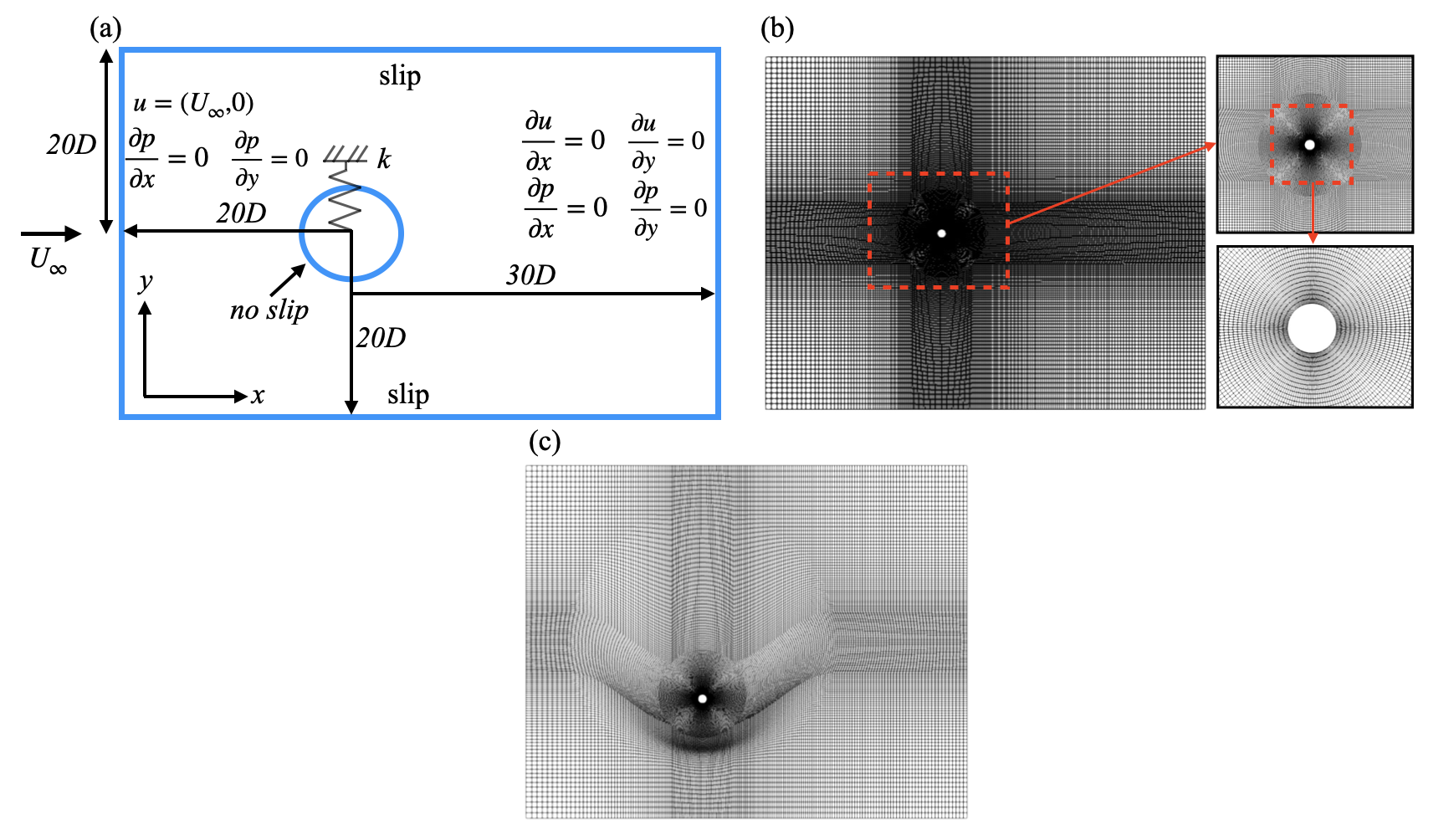}
    \caption{(a) The CFD problem domain and boundary conditions, (b) the fluid mesh, and (c) the fluid mesh during mesh deformation.}
    \label{figure:  CFD domains and mesh}
\end{figure}

The computational domain and boundary conditions implemented in {\tt OpenFOAM} are illustrated in Fig.~\ref{figure:  CFD domains and mesh}(a). From the centre of the cylinder $(0,0)$, the domain extends $30D$ downstream, $20D$ upstream and $20D$ above and below. At the inlet, a uniform free-stream velocity $U_\infty$ and a zero-gradient pressure boundary condition (BC) are applied. At the outlet, zero-gradient BCs are applied for both velocity and pressure. Slip BCs are applied along the top and bottom walls with a zero-gradient pressure BC. The moving cylinder is assigned a moving wall velocity BC and a zero-gradient pressure BC. The Arbitrary Lagrangian-Eulerian approach is adopted, using a morphing-mesh formulation in which a single deforming mesh is employed. The mesh is shown in Fig.~\ref{figure:  CFD domains and mesh}(b), along with a close-up view around the cylinder. The mesh is locally refined around the cylinder to accommodate large deformations whilst maintaining cell connectivity and avoiding negative cell volume errors; see Fig.~\ref{figure:  CFD domains and mesh}(c).

\begin{figure}[htbp]
    \centering
    \includegraphics[width=1\textwidth]{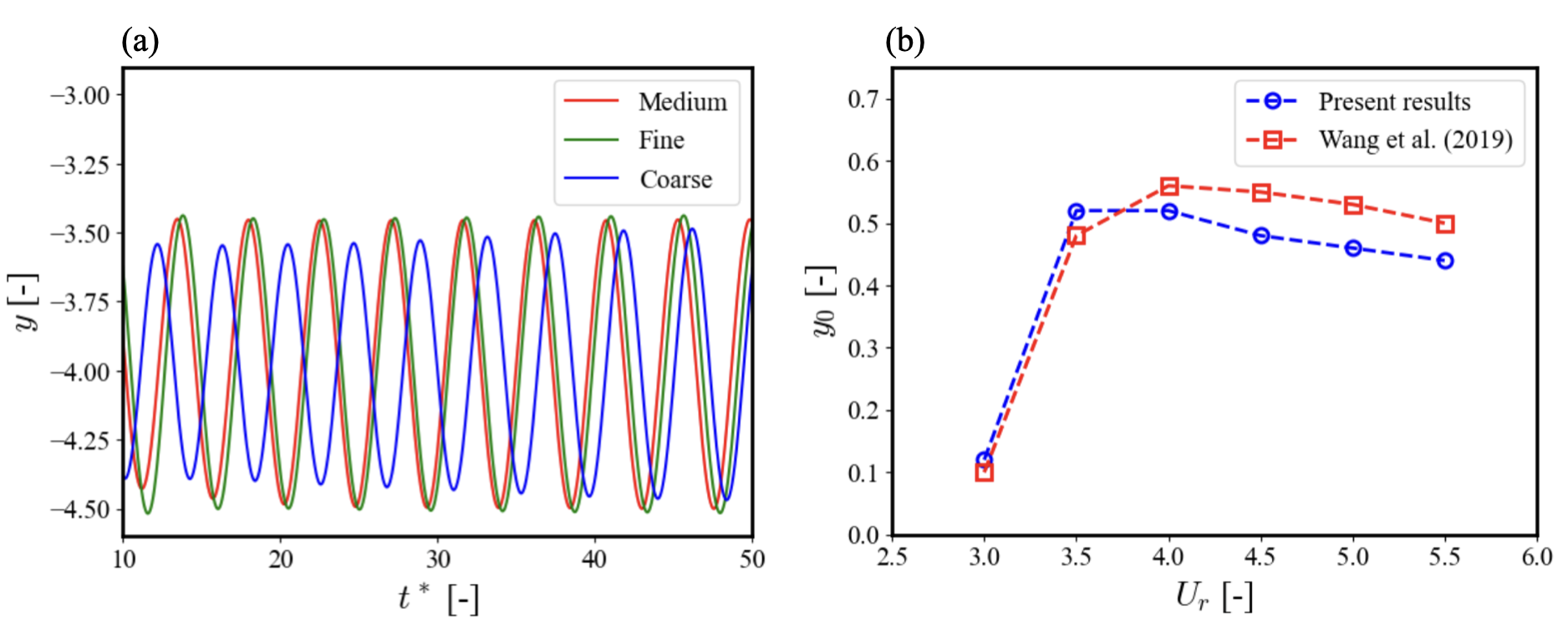}
    \caption{(a) A comparison of the non-dimensional structural displacement for medium, fine and coarse meshes between the interval $t^* = 10$ to $t^* = 50$. (b) The dimensionless amplitude of oscillation $y_0$ plotted against $U_r$ and validated with the results of \citet{wang2017three}.}
    \label{figure: verification and validation}
\end{figure}

\begin{table}[htbp]
    \centering
    \caption{Relative error between the maximum displacement results obtained from the three tested meshes.}
    \setlength{\tabcolsep}{8pt} 
    \renewcommand{\arraystretch}{1.3} 
    \begin{tabular}{l c c}
        \hline
        Mesh   & $y_0$ & Relative error (\%) \\ 
        \hline
        Fine   & 0.539 & 2.783\\
        Medium & 0.524 &  6.298 \\
        Coarse & 0.491 & -- \\
        \hline
    \end{tabular}
    \label{relative-error-between-meshes}
\end{table}


The optimal grid size is determined through a grid independence study by comparing results from fine, medium, and coarse meshes with cell counts of 136168, 73208, and 27688, respectively. The temporal evolution of the dimensionless transverse displacement ($y=Y/D$) was compared across the three meshes for a case with $U_r=4$ over dimensionless time $ t^*$ = $tU_\infty/D$, as shown in Fig.~\ref{figure: verification and validation}(a). The fine- and medium-mesh results exhibit close agreement, whereas the coarse-mesh results deviate significantly. Table \ref{relative-error-between-meshes} presents the relative error in the maximum $y$ values between the three meshes. The error between the fine and medium meshes ($2.783 \%$) is significantly smaller than between the medium and coarse meshes ($6.298 \%$). Based on these observations, the medium mesh is deemed sufficient for the present study, as it offers acceptable accuracy and computational efficiency. A fixed time-step size of $\delta t = 0.0001$ is used, ensuring that the Courant–Friedrichs–Lewy (CFL) number remains consistently below 1 throughout the simulations. The present simulation results are validated against those of \citet{wang2017three} by comparing the dimensionless oscillation amplitude $y_0$ for varying $U_r$. It is seen that the present simulation results closely agree with the amplitude response plot presented in the reference paper, thereby validating the present computational methodology; see Fig.~\ref{figure: verification and validation}(b). It is worth noting that the deviation in $y_0$ values may be attributed to differences in computational methodology, solver setup, and mesh density used in this study and those of \citet{wang2017three}.

\section{Proper orthogonal decomposition analysis}
\label{app:C}

\noindent Proper Orthogonal Decomposition (POD), derived from principal component analysis \citep{berkooz1993proper, holmes1996turbulence}, is used to extract dominant coherent structures from complex flow data, providing a low-dimensional representation of a high-fidelity system. In this study, we use a snapshot-based POD approach \citep{sirovich1987turbulence}, in which the simulation results are processed as a collection of snapshots of the flow field $u(x,t)$ at different time instants. These snapshots are organised into a flattened matrix $U$, where $U \in \mathbb{R}^{m \times n}$, $m$ represents the number of spatial points and $n$ the number of time samples. Here, the $n$th column contains all coordinates at the $n$th timestep. The next step is to apply Singular Value Decomposition (SVD) on the $U$ matrix to decompose it as:

\begin{equation}
U = \Phi \Sigma V^T,
\end{equation}

\noindent where $\Phi \in \mathbb{R}^{m \times m}$ contains spatial modes orthogonal in space and time, $\Sigma \in \mathbb{R}^{m \times n}$ is a diagonal matrix of singular values, and $V \in \mathbb{R}^{n \times n}$ contains the temporal coefficients. The spatial modes $\phi_i(x)$ are the columns of $\Phi$, and the rows of $\Sigma V^T$ represent the temporal dynamics. The POD decomposition expresses the flow field as a linear combination of these modes:

\begin{equation}
u(x, t) = \sum_{i=1}^{N} a_i(t) \phi_i(x).
\end{equation}

\noindent Here, $a_i(t)$ are the temporal coefficients and $\phi_i(x)$ are the spatial modes. The modes $\phi_i(x)$ are arranged in descending order of their singular values $\sigma_i$, which reflect the energy content of each mode. The cumulative energy captured by the first $k$ modes is given by:

\begin{equation}
E_k = \frac{\sum_{i=1}^{k} \sigma_i^2}{\sum_{i=1}^{N} \sigma_i^2},
\end{equation}

\noindent where $N$ is the total number of modes. To reduce the system's dimensionality, truncation is applied by selecting the first $k$ modes according to their cumulative energy. The system can then be represented in terms of these $k$ modes, with the $ a_i (t) $ as basis functions that form a new coordinate system, where $a_i(t)$ describe the temporal dynamics of each mode. This transformation enables the projection of high-dimensional data onto a lower-dimensional subspace, thereby significantly reducing computational cost by providing an efficient basis for the system. This technique is a preliminary step towards applying SINDy to high-dimensional systems, such as those encountered in CFD. Without projecting the dynamics onto a low-dimensional subspace, SINDy struggles to identify stable, interpretable dynamics. POD improves computational efficiency while enabling SINDy to effectively discover governing equations in a reduced-dimensional space that captures the system's dominant dynamics.

\begin{figure}[htbp]
    \centering
    \includegraphics[width=1\textwidth]{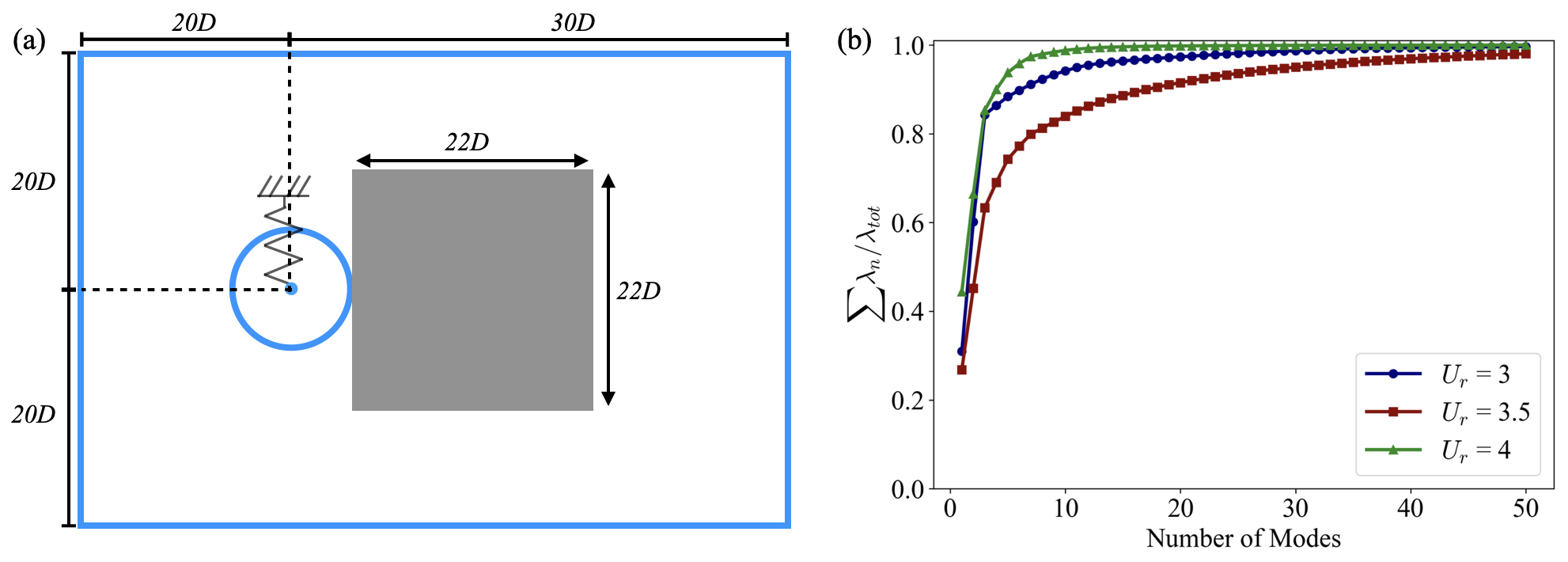}
    \caption{(a) The spatial domain selected for the POD analysis, restricted to a near-wake subdomain extending $22D$ downstream and $22D$ in the cross-flow direction (not drawn to scale). (b) Cumulative modal energy content $\sum \lambda_n / \lambda_\mathrm{tot}$ as a function of the number of retained POD modes for $U_r = 3.0$, $U_r = 3.5$, and $U_r = 4.0$, illustrating the rate of energy convergence across the three reduced velocities and the more diffuse modal energy distribution characteristic of the transitional pre-lock-in dynamics at $U_r = 3.5$.}
    \label{fig:pod_energy}
\end{figure}

POD is applied to the near-wake velocity field data obtained from the CFD simulations across three representative reduced velocities, $U_r \in \{3.0,\, 3.5,\, 4.0\}$, encompassing both the aperiodic pre-lock-in and periodic lock-in regimes, to assess the feasibility of using SINDy to reconstruct the dominant spatiotemporal flow dynamics from a low-dimensional modal representation. Figure~\ref{fig:pod_energy}(a) illustrates the spatial domain selected for the POD analysis, which is restricted to a near-wake subdomain extending $22D$ downstream and $22D$ in the cross-flow direction, chosen to capture the dominant vortex formation and shedding dynamics whilst reducing the dimensionality of the snapshot matrix and the computational cost of the decomposition.

Figure~\ref{fig:pod_energy}(b) presents the cumulative modal energy content $\sum \lambda_n / \lambda_\mathrm{tot}$ as a function of the number of retained POD modes for each reduced velocity. At $U_r = 4.0$, the energy distribution is highly concentrated in the leading modes, with the first ten modes capturing in excess of 90\% of the total flow energy, consistent with the well-organised, single-frequency periodic wake topology observed in the FTLE visualisations at this reduced velocity (Fig.~\ref{figure:FTLE}). A similarly rapid energy convergence is observed at $U_r = 3.0$, where the first ten modes again exceed the 90\% threshold. At $U_r = 3.5$, however, the modal energy distribution is markedly more diffuse, with ten modes capturing considerably less cumulative energy than at the other two reduced velocities, reflecting the broader spectral content and less organised vortical structures characteristic of the transitional pre-lock-in dynamics at this reduced velocity. Based on these observations, ten POD modes are retained for all three cases to maintain consistency across the identification campaign and to provide a sufficiently reduced-order basis for capturing the dominant coherent structures whilst limiting the dimensionality of the SINDy state vector. The flow-field snapshots used for the POD analysis are extracted over the dimensionless time interval $t^* \in [50, 150]$ at uniform intervals of $\Delta t^* = 0.1$, yielding a total of 1000 snapshots per case, and the corresponding scaled temporal coefficients of the retained modes are subsequently used as input to the SINDy identification algorithm.

It is important to note that retaining 10 POD modes as state variables yields a relatively high-dimensional identification problem compared to the two-variable wake-oscillator formulation in Sec.~\ref{sec:4}. The lower-energy modes, which individually contribute little to the total energy but collectively represent residual wake interactions and transitional dynamics, exhibit irregular temporal behaviour that poses significant challenges for stable sparse model identification, particularly when higher-order polynomial libraries are employed to capture nonlinear inter-modal interactions.

\section{Representative example of the identified low-order models at $U_r = 5$}
\label{app:D}

\noindent This appendix presents the governing equations discovered by each identification approach at a representative reduced velocity of $U_r = 5$, which lies within the lock-in regime and is therefore characterised by large-amplitude, strongly periodic structural oscillations. Four sets of equations are presented: the ground-truth reduced mathematical model obtained from the coupled wake oscillator, the equations identified by standard SINDy applied to synthetic data generated from that model, and the equations identified by standard SINDy and WSINDy applied directly to CFD data. Comparison of these four sets illustrates how the complexity, sparsity, and physical interpretability of the identified models vary across identification approaches and data sources, and serves as a concrete example of the broader trends discussed in the main body of the paper.

\begin{description}
\label{equations_Ur5}
\item[\textbf{Reduced mathematical model:}]
\begin{align*}
\dot{y}_1 &= y_2, \\
\dot{y}_2 &= 0.04188\, q_1 - 0.6771\, y_2 - y_1, \\
\dot{q}_1 &= q_2, \\
\dot{q}_2 &= 12.0(0.04188\, q_1 - 0.6771\, y_2 - y_1) - 0.3(q_1^2 - 1)q_2 - q_1.
\end{align*}

\item[\textbf{SINDy-predicted model from numerical intergartion data:}]
\begin{align*}
\dot{y}_1 &= y_2, \\
\dot{y}_2 &= -1.024\, y_1 - 0.677\, y_2 + 0.0683\, q_1, \\
\dot{q}_1 &= q_2, \\
\dot{q}_2 &= -12.37\, y_1 - 4.87\, y_2 - 0.89\, q_1 + 0.300\, q_2 - 0.38\, q_1^2 q_2.
\end{align*}

\item[\textbf{SINDy-predicted model from CFD data:}]
\begin{align*}
\dot{y}_1 &= y_2, \\
\dot{y}_2 &= -4.060 + 1.009\, y_1 + 0.031\, q_1 + 0.20\, y_1^2 + 0.15\, y_2^2 ,\\
\dot{q}_1 &= 121.276 + 41.286\, y_1 + 0.085\, y_2 + 1.915\, q_1 + 0.132\, q_2 \\
&\quad + 3.490\, y_1^2 + 0.459\, y_1 q_1 - 0.144\, y_1 q_2 + 2.262\, y_2^2 \\
&\quad + 0.075\, y_2 q_1 + 0.042\, y_2 q_2 + 0.054\, q_1^2 - 0.029\, q_1 q_2, \\
\dot{q}_2 &= 2060.594 + 693.370\, y_1 - 50.928\, y_2 + 62.342\, q_1 - 9.617\, q_2 \\
&\quad + 58.034\, y_1^2 - 8.604\, y_1 y_2 + 12.061\, y_1 q_1 - 1.699\, y_1 q_2 \\
&\quad + 33.154\, y_2^2 - 2.668\, y_2 q_1 + 1.731\, y_2 q_2 + 1.674\, q_1^2 \\
&\quad - 0.650\, q_1 q_2 + 0.113\, q_2^2.
\end{align*}

\item[\textbf{WSINDy-predicted model from CFD data:}]
\begin{align*}
\dot{y}_1 &= y_2, \\
\dot{y}_2 &= 1.331\, y_1 + 0.008\, q_1 + 0.213\, y_1^2 + 0.159\, y_2^2, \\
\dot{q}_1 &= -0.078\, y_1 - 0.069\, q_1 + 0.998\, q_2 - 0.013\, y_1^2 + 0.003\, y_1 y_2, \\
\dot{q}_2 &= -2.331\, y_1 - 2.644\, y_2 - 0.424\, q_1 - 0.038\, q_2 \\
&\quad - 0.372\, y_1^2 - 0.268\, y_2^2 - 0.364\, y_1 y_2 + 0.067\, q_1 q_2.
\end{align*}

\end{description}

\section*{Acknowledgements}
\noindent The authors gratefully acknowledge the support of the UK Engineering and Physical Sciences Research Council (EPSRC) for the DLA scholarship of Hibah Saddal, grant EP/W524396/1. The computations described in this paper were performed on the University of Birmingham's BlueBEAR HPC service, which provides high-performance computing to the University's research community. See \url{http://www.birmingham.ac.uk/bear} for more details.

\section*{CRediT authorship contribution statement}
\noindent Haimi Jha: Data curation, Methodology, Validation, Formal Analysis, Writing-original draft. Hibah Saddal: Simulations, Data curation, Methodology, Validation, Writing- review. Chandan Bose: Conceptualisation, Simulation, Data curation, Methodology, Formal Analysis, Writing- original
draft, review \& finalization.
	
\section*{Declaration of interests}
\noindent The authors report no conflict of interest.

\bibliographystyle{elsarticle-harv} 
\bibliography{bibliography}






\end{document}